  \documentclass{aa}  

\usepackage{multirow}
\usepackage{xcolor}
\newcommand\crule[3][black]{\textcolor{#1}{\rule{#2}{#3}}}
\definecolor{tab:blue}{RGB}{31,119,180}
\definecolor{tab:cyan}{RGB}{23,190,207}
\definecolor{tab:red}{RGB}{214,39,40}
\definecolor{tab:orange}{RGB}{255,127,14}
\definecolor{tab:green}{RGB}{44,160,44}
\definecolor{tab:olive}{RGB}{188,189,34}
\definecolor{tab:purple}{RGB}{148,103,189}
\definecolor{tab:pink}{RGB}{227,119,194}
\usepackage{graphicx}
\usepackage{txfonts}
\usepackage{float}

\usepackage[version=4]{mhchem}
\usepackage[colorlinks=true,
            linkcolor=red,
            urlcolor=blue, citecolor=blue]{hyperref}
\usepackage{subfig}

\begin{document}

   \title{Large Interferometer For Exoplanets (LIFE):}

   \subtitle{V. Diagnostic potential of a mid-infrared space-interferometer for studying Earth analogs } 
   \titlerunning{\textit{LIFE}: V. Diagnostic potential of a mid-infrared space-interferometer for studying Earth analogs}

   \author{Eleonora Alei \inst{1, 2}\fnmsep\thanks{Correspondence: \texttt{elalei@phys.ethz.ch}}, Björn S. Konrad\inst{1, 2},   Daniel Angerhausen\inst{1, 2, 3}, John Lee Grenfell\inst{4},  Paul Mollière\inst{5}, Sascha P. Quanz\inst{1, 2}, Sarah Rugheimer\inst{6}, Fabian Wunderlich\inst{4}, and the \emph{LIFE} collaboration\thanks{Website:  \url{www.life-space-mission.com}}}
   \institute{
    ETH Zurich, Institute for Particle Physics \& Astrophysics, Wolfgang-Pauli-Str. 27, 8093 Zurich, Switzerland
    \and
    National Center of Competence in Research PlanetS (www.nccr-planets.ch)
    \and Blue Marble Space Institute of Science, Seattle, United States
    \and Department of Extrasolar Planets and Atmospheres (EPA), Institute for Planetary Research (PF), German Aerospace Centre (DLR), Rutherfordstr. 2, 12489 Berlin
    \and  Max-Planck-Institut f\"ur Astronomie, Königstuhl 17, 69117 Heidelberg, Germany
    \and Department of Physics, University of Oxford, Oxford, OX1 3PU, UK }
    \authorrunning{Alei, E. et al.}
   \date{Received ; accepted }

 
  \abstract
   {An important future goal in exoplanetology is to detect and characterize potentially habitable planets.  Concepts for future space missions have already been proposed: from a large UV-Optical-Infrared space mission for studies in reflected light, to the Large Interferometer for Exoplanets (\emph{LIFE}) for analyzing the thermal portion of the planetary spectrum. Using nulling interferometry, \emph{LIFE} will allow us to constrain the radius and effective temperature of (terrestrial)  exoplanets, as well as provide unique information about their atmospheric structure and composition. }
   {We explore the potential of \emph{LIFE} in characterizing emission spectra of Earth at various stages of its evolution. This allows us (1) to test the robustness of Bayesian atmospheric retrieval frameworks when branching out from a Modern Earth scenario while still remaining in the realm of habitable (and inhabited) exoplanets, and (2)  to refine the science  requirements for \emph{LIFE} for the detection and characterization of habitable, terrestrial exoplanets. }
   {We perform Bayesian retrievals on simulated spectra of 8 different scenarios, which correspond to cloud-free and cloudy spectra of four different epochs of the evolution of the Earth. Assuming a distance of 10 pc and a Sun-like host star, we simulate observations obtained with \emph{LIFE} using its simulator LIFE\textsc{sim}, considering all major astrophysical noise sources.}
   {With the nominal spectral resolution (R $=50$) and signal-to-noise ratio (assumed to be S/N~$=10$ at 11.2~$\mu$m), we can identify the main spectral features of all the analyzed scenarios (most notably \ce{CO2}, \ce{H2O}, \ce{O3}, \ce{CH4}). This allows us to distinguish between inhabited and lifeless scenarios. Results suggest that particularly \ce{O3} and \ce{CH4} yield an improved abundance estimate by doubling the S/N from 10 to 20. 
 Neglecting clouds in the retrieval still allows for a correct characterization of the atmospheric composition. However, correct cloud modeling is necessary to avoid biases in the retrieval of the correct thermal structure.}
   {From this analysis, we conclude that the baseline requirements for R and S/N are sufficient for \emph{LIFE} to detect \ce{O3} and \ce{CH4} in the atmosphere of an Earth-like planet with an abundance of \ce{O2} of around 2\% in volume mixing ratio. Doubling the S/N would allow a clearer detection of these species at lower abundances. This information is relevant in terms of the \emph{LIFE} mission planning. We also conclude that cloud-free retrievals of cloudy planets can be used to characterize the atmospheric composition of terrestrial habitable planets, but not the thermal structure of the atmosphere. From the inter-model comparison performed, we deduce that differences in the opacity tables (caused by e.g. a different line wing treatment) may be an important source of systematic errors.}

   \keywords{    Methods: statistical --
                Planets and satellites: terrestrial planets --
                Planets and satellites: atmospheres
               }

   \maketitle
%

\section{Introduction}\label{sec:intro}

Temperate terrestrial exoplanets are predicted to be very abundant in our galaxy \citep{Bryson2021}. These planets are ideal candidates when searching for life beyond our Solar System. 
A powerful way to characterize a terrestrial exoplanet in the context of its habitability is by detecting and studying its atmosphere with the goal to constrain its surface conditions. Atmospheric spectra are influenced by many parameter and processes, such as the chemical composition, the temperature structure of the atmosphere, the presence of clouds, as well as  emission and scattering from the surface.

The detection and characterization of potentially habitable, rocky exoplanets is challenging with current facilities. 
For this reason, there is a widespread interest in the community to build new instruments for the search of life in the universe, as reported in the White Paper series in the context of the ESA ``Voyage 2050'' process\footnote{\url{https://www.cosmos.esa.int/web/voyage-2050}}, as well as the US Astro 2020 Decadal survey \citep{NAP26141}.
Space missions that aim at characterizing terrestrial exoplanets have been proposed, such as HabEx \citep{Gaudi2020} and LUVOIR \citep{Peterson2017} focusing on the reflected (visible and near-infrared) portion of the planetary spectrum, as well as \emph{LIFE} \citep[Large Interferometer for Exoplanets,][hereafter Paper I]{Quanz2021}, which will characterize terrestrial planets in the thermal (mid-infrared), emitted portion of the planetary spectrum. 
Using nulling interferometry, \emph{LIFE} will allow us to constrain the radius and effective temperature of (terrestrial)  exoplanets, as well as provide unique information about their atmospheric structure and composition \citep[][hereafter Paper II and III, respectively]{2022arXiv220300471D,konrad2021large}. 

Due to the current lack of high-quality observational data, we must rely momentarily on simulated observations of terrestrial planets to create and improve the analyses algorithms, but also to provide scientific and technical requirements when planning a mission. 
This  effort is currently ongoing within the \emph{LIFE} Initiative and  in a previous study (Paper III), we built a Bayesian retrieval routine to estimate the planetary and atmospheric parameters of a simulated Modern Earth twin at 10 pc distance as it would be observed by  \emph{LIFE}. In this work, we extend this exercise to other stages in the evolution of Earth's atmosphere. 

Our planet has been habitable for about 4.4 billion years \citep[see e.g.][and references therein]{Heller2021}. In this context, we define a planet as habitable if its physical and chemical conditions would allow water, if present, to be liquid on the surface.

In the prebiotic stage of Earth's evolution, the atmosphere lacked \ce{O2} (currently about 21\% of the atmospheric composition by volume). It was instead  a \ce{CO2}-\ce{N2}-\ce{H2O}-rich atmosphere, with traces of \ce{CH4} from volcanism. The early forms of life developed under a reducing environment and survived under anaerobic conditions \citep[and references therein]{Olson2018}. Methanogenesis was thought to be a dominant metabolism at this stage (around 3.5 Ga), which would explain the increase in \ce{CH4} in the atmosphere \citep[see e.g.][]{Wolfe2018}. 

Around 3 Ga, life forms that could use carbon dioxide to produce oxygen (via oxygenic photosynthesis) appeared \citep{doi:10.1126/science.289.5485.1703}. These eventually led to a significant increase of \ce{O2} maximally up to $\sim 1\%$ PAL\footnote{Present Atmospheric Level} \citep[see][and references therein]{2021E&PSL.56116818G,Lyons2014,Lyons2021} around 2.33 Ga \citep{doi:10.1126/sciadv.1600134}, during the so-called "Great Oxygenation Event" (GOE). There is also evidence pointing to a second increase in the \ce{O2} abundance (up to $\sim10\%$ PAL) occurred around 0.8 Ga, in the "Neoproterozoic Oxygenation Event" (NOE) \citep{ShieldsZhou2011,2010GeCoA..74.4187C}.

The high abundance of carbon dioxide in the early Earth would have enhanced the atmospheric greenhouse effect, allowing Earth to be habitable despite the fainter solar irradiation \citep[see e.g.][and references therein]{2012RvGeo..50.2006F}. The positive feedback between the carbon-silicate cycle and the increase in irradiation would have then allowed to maintain temperatures conducent to liquid water over the last 4 Ga. The increase in irradiation from the Sun over the eons has made the weathering of \ce{CO2} more efficient, decreasing the amount of carbon dioxide in the atmosphere and thus dampening the atmospheric greenhouse effect \citep[see e.g.][and references therein]{doi:10.1089/ast.2020.2411}. The appearance of photosynthetic life forms and the onset of plate tectonics also contributed to the depletion of atmospheric \ce{CO2}.

Numerous processes including biology and geology have driven the wide-ranging evolution of Earth's atmosphere during the various epochs of its development. Our modern atmosphere represents however only a small fraction of Earth's evolutionary states. It is therefore important to simulate a suitable range of different atmospheric epochs from Earth's history when investigating Earth-like atmospheres.
For this study, we simulated observations obtained by \emph{LIFE} starting from theoretical spectra of 4 distinct epochs of Earth's atmospheric evolution, produced from a self-consistent 1D climate and photochemistry model coupled with a line-by-line radiative transfer model \citep{Rugheimer2018}.  The observed spectra were simulated using the \emph{LIFE} noise simulator LIFE\textsc{sim} (for details on the simulator see Paper II). We then used the Bayesian retrieval routine presented in Paper III to characterize the different atmospheres. 

We aim to address the following research questions:
\begin{itemize}
    \item \emph{science-driven questions}: How well could \emph{LIFE} characterize atmospheres of habitable planets? Could \emph{LIFE} differentiate between different atmospheres, and with what confidence? What is the impact of clouds on this assessment?  What are the most promising (combinations of) detectable biosignatures? 
    \item \emph{technology- and computationally-driven questions}: Is the combination of spectral resolution ($R=\lambda/\Delta\lambda$), signal to noise ratio (S/N) and wavelength range defined in Paper III still adequate for this case study? What are the caveats and limitations of the Bayesian retrieval routine? What systematics may arise when comparing two different models (e.g. in terms of differences in line lists, scattering treatment, identification of biomarkers)? 
\end{itemize}

We discuss how we adapted the input spectra to simulate \emph{LIFE} observations and describe the grid of scenarios in Section~\ref{sec:methods}. We show and describe the results in Section~\ref{sec:results}. A  thorough discussion of our findings and of the potential systematic uncertainties of the retrieval routine is provided in Section~\ref{sec:discussion}. In Section~\ref{sec:conclusions} we report the main takeaway points from this study, and in Section \ref{sec:future} we trace an outlook of the ongoing and future studies.

\section{Methods}\label{sec:methods}

We start by discussing the details of the input spectra that were used in this study (Section~\ref{sec:input_spectra}).  We then discuss the updates on the Bayesian retrieval routine (Section~\ref{sec:framework}). We describe the assumptions that our model takes into account and the main potential source of systematic errors in the retrievals in Section~\ref{sec:limitations}.

\subsection{Input spectra and scenarios}\label{sec:input_spectra}

We consider spectra corresponding to four different evolutionary epochs of Earth: the prebiotic Earth (3.9 Ga), the Earth shortly after the Great Oxygenation Event (2.0 Ga), the Neoproterozoic Oxygenation Event (0.8 Ga), and Modern Earth. All considered Earth spectra were produced by \citet{Rugheimer2018}. These self-consistent spectra were produced using a 1D convective-radiative transfer model loosely coupled with a 1D climate model and a 1D photochemistry model. The authors accounted for the thermal chemistry and the photochemistry of more than 55 species. The atmospheres were modeled up to $10^{-4}$ bar and split into 100 layers. The radiative forcing of clouds is included by adjusting the surface albedo of the planet.

The results of the photochemistry-climate-radiative model were then fed to a line-by-line radiative transfer model to produce emission spectra.  The line lists and the pressure broadening coefficients were from to the HITRAN 2016 database \citep{2017JQSRT.203....3G}. Surface scattering was included in the calculations, assuming 70\% ocean, 2\% coast, and 28\% land. 
In some scenarios, a partial cloud coverage was directly included in the calculation of the emission spectrum. In these cloudy cases, the authors assumed a 60\% cloud coverage (split into 40\% water clouds at 1 km altitude, 40\% water clouds at 6 km altitude, and 20\% ice clouds at 12 km altitude) consistent with an averaged Earth cloud model. Aerosol was not included in the calculation. 
The model has been validated from the visual to the mid-infrared wavelength ranges with observations of Earth \citep{Kaltenegger2007,KalteneggerTraub2009,Rugheimer2013}. For further details, we refer the reader to \citet{Rugheimer2018}.

\begin{table*}[]
\renewcommand{\arraystretch}{1.5}

\caption{Model description, identifiers, and colors. }             
\label{table:id}      
\centering                          
\begin{tabular}{rcl}    
\hline\hline                 
Identifier & Color & Model Description\\
\hline

MOD-CF & \ \crule[tab:blue]{0.7cm}{0.3cm} & Modern Earth, cloud-free sky \\
 MOD-C & \ \crule[tab:cyan]{0.7cm}{0.3cm} & Modern Earth, cloudy sky\\
NOE-CF & \ \crule[tab:red]{0.7cm}{0.3cm} & Neoproterozoic Oxygenation Event Earth, cloud-free sky\\
 NOE-C & \ \crule[tab:orange]{0.7cm}{0.3cm} & Neoproterozoic Oxygenation Event Earth, cloudy sky\\
 GOE-CF & \ \crule[tab:green]{0.7cm}{0.3cm} & Great Oxygenation Event Earth, cloud-free sky\\
  GOE-C & \ \crule[tab:olive]{0.7cm}{0.3cm} & Great Oxygenation Event Earth, cloudy sky\\
 PRE-CF & \ \crule[tab:purple]{0.7cm}{0.3cm} & Prebiotic Earth, cloud-free sky\\
 PRE-C & \ \crule[tab:pink]{0.7cm}{0.3cm} & Prebiotic Earth, cloudy sky\\

\hline 
\end{tabular}
\tablefoot{The identifier "CF" denotes "cloud-free" (meaning "clear sky"). The identifier "C" denotes "cloudy". See main text for details. }
\end{table*}

For each epoch, we consider both clear sky and cloudy sky spectra, which yields a total of 8 scenarios. 
We assigned every modeled scenario an identifier and a specific color, as listed in Table~\ref{table:id}. We will use these identifiers throughout the remainder of the paper.

We simulate observations with \emph{LIFE} via the LIFE\textsc{sim} tool (see Paper II for a description of the simulator). LIFE\textsc{sim} estimates the wavelength-dependent S/N considering all major astrophysical noise sources (stellar leakage, local zodiacal dust emission, and exo-zodiacal dust emission).  We consider an Earth-sized planet on a 1 AU orbit around a Sun-like star at 10 pc distance. For our baseline analyses we assumed the nominal simulation parameters for LIFE\textsc{sim} as summarized in Table~\ref{tab:lifesim} (see Paper I and Paper II for details). We consider an exo-zodi level of 3 times the local zodiacal dust density, based on the results from the HOSTS survey \citep{2020AJ....159..177E}.

\begin{table}[]
\caption{Simulation parameters used in LIFE\textsc{sim} for the baseline analyses.}              
\label{tab:lifesim}      
\centering                                      
\begin{tabular}{l l}          
\hline\hline                        
Parameter & Value \\    
\hline                                   
    Detector quantum efficiency & 0.7\\      
    Total instrument throughput & 0.05\\
    Minimum Wavelength & 4~$\mu$m       \\
    Maximum Wavelength & 18.5~$\mu$m   \\
    Spectral Resolution & 50      \\
    S/N\tablefootmark{a}  & 10\\
    Interferometric Baseline & 10-100~m \\
    Aperture Diameter & 2~m \\
    Exozodi level & 3 $\times$ local zodiacal dust \\
    Planet radius & 1~${R_\oplus}$\\
    Distance to the system & 10~pc\\
\hline
\end{tabular}
\tablefoot{See Paper I and Paper II for details.\tablefoottext{a}{This S/N is fixed at a wavelength of 11.2~$\mu$m and the S/R of all other spectral bins is computed via LIFE\textsc{sim}.}}
\end{table}

\subsection{Updates on the Bayesian retrieval framework}\label{sec:framework} 

We denote the input spectra from Section~\ref{sec:input_spectra} as the "true spectra". To simulate a  \emph{LIFE}-like observation of these targets, we run LIFE\textsc{sim} on the true spectra, thus obtaining simulated "observed spectra".  
We then perform a retrieval on the observed spectra using \texttt{petitRADTRANS} \citep{2019A...627A..67M}  as "forward model" in the retrieval routine, and the Bayesian sampler model \texttt{pyMultiNest} \citep{Buchner:PyMultinest} as "parameter estimation routine" (cf. Paper III).

The theoretical 1D atmospheric model \texttt{petitRADTRANS} \citep{2019A...627A..67M} applies the radiative transfer equation to calculate spectra corresponding to a set of parameters. These parameters describe the bulk parameters (planetary mass and radius), the pressure-temperature (P-T) structure (approximated by a fourth-order polynomial), and the chemical composition of the atmosphere. 

Our Bayesian retrieval framework recursively draws combinations of parameters from a set of "priors" that describe the "a priori" probability distribution of each parameter (listed in Table~\ref{table:parameters}) and uses the forward model to compute the corresponding spectra. Then, the Bayesian framework tests how well these calculated spectra fit the observed one using a "likelihood" function (see Eq. (3) in Paper III). In order to sample the prior space efficiently, our retrieval relies on the parameter estimation routine \texttt{pyMultiNest} \citep{Buchner:PyMultinest}, which is based on \texttt{MultiNest} \citep{2009MNRAS.398.1601F}. This routine applies the Nested Sampling algorithm \citep{Skilling:Nested_Sampling} to fit the theoretical spectral model to the observed spectrum, and thereby yields estimates and uncertainties for the model parameters.
These estimates are the "posterior probability distributions" (or "posteriors"). The posteriors contain the information on which combinations of model parameters best describe the observed spectrum. For more details about the Bayesian retrieval framework, we refer the reader to Paper III.

In our previous work, we argued that effects of scattering on simulated mid-infrared spectra are negligible at the considered resolutions and LIFE\textsc{sim} noise patterns. In that study, we had to be particularly mindful of the computing time. The version of \texttt{petitRADTRANS} used in Paper III only allowed to calculate spectra at R~$=1000$, which were subsequently binned down to the resolution of the input spectrum (R~$=35 - 100$ in Paper III). Calculating a spectrum at R~$=1000$ excluding scattering required $\approx0.5$ seconds, whereas including scattering required $\approx18$ seconds. Since millions of spectra have to be calculated for a single Bayesian retrieval, including scattering was prohibitive with respect to the computing time, especially when considering a large grid of retrieval runs. 
Recent updates to \texttt{petitRADTRANS} have enabled us to compute spectra at any resolution, provided a grid of correlated-k tables at that resolution is available. These tables can be produced with \texttt{petitRADTRANS} by binning down the R~$=1000$ opacity tables. We compiled a correlated-k opacity database for R~$= 50$ and used it to produce spectra directly at this resolution. This reduces the computation time per spectrum significantly and allows us to compute emission spectra in $\approx0.04$ seconds when excluding scattering, and in $\approx0.5$ seconds when including scattering. This reduction in computing time allows us to include scattering in the theoretical spectral model. 

Other updates on \texttt{petitRADTRANS} were performed. The treatment of collision-induced absorption (CIA) was modified by updating the interpolation of the CIA tables, originally in FORTRAN, to Python. Minor variations in the interpolation options (log-linear compared to nearest neighbor in Paper III) make the new model not directly comparable with its previous version. However, at Earth-like conditions, the differences in the CIA signature on the spectrum remain negligible. Also, the CIA features impact mostly the short wavelengths ($\lambda<6\ \mu m$), where the LIFE\textsc{sim} noise is large. Furthermore, the treatment of scattering was updated. Originally implemented in \texttt{petitRADTRANS} only for gaseous planets \citep[see][]{2019A...627A..67M}, it was adapted to include the scattering by a rocky surface. More information on the new implementation of scattering can be found in Appendix~\ref{app:scattering}, as well as the \texttt{petitRADTRANS} documentation\footnote{\url{https://petitradtrans.readthedocs.io}}. This new version of \texttt{petitRADTRANS} is available in the main GitLab repository\footnote{\url{https://gitlab.com/mauricemolli/petitRADTRANS}}.

Using the updated retrieval framework, we run retrievals for the eight different spectra introduced in Section~\ref{sec:input_spectra}. We retrieve the same parameters as in Paper III and leave most prior distributions unchanged. An exhaustive list of the parameters and priors used, and the corresponding expected values for the different epochs, are listed in Table~\ref{table:parameters}. Most priors are represented by a boxcar function (hereafter "uniform" priors): every value is equally probable if within a certain range. Two notable exceptions are the priors for the planetary radius ${R_\mathrm{pl}}$ and the logarithm of the planetary mass ${\log_{10}}(M_\mathrm{pl})$, whose priors are Gaussian. The prior we assumed on ${R_\mathrm{pl}}$ is based on the radius estimate expected from observing a terrestrial planet with \emph{LIFE} during its search phase (see Paper II for details). The ${\log_{10}}(M_\mathrm{pl})$ prior was inferred from  ${R_\mathrm{pl}}$  using the statistical mass-radius relation presented in \citet{Kipping:Forecaster} (see Paper III for details).

\begin{table*}[!htbp]
\renewcommand{\arraystretch}{1.5}

\caption{Summary of the parameters used in the retrievals, their expected values and their prior distributions. }             
\label{table:parameters}      
\centering                          
\begin{tabular}{lllccccc}    
\hline\hline                 
\multirow{2}{*}{Parameter} &\multirow{2}{*}{Description}  &\multirow{2}{*}{Prior} &  \multicolumn{4}{c}{Expected values \citep[][]{Rugheimer2018}}\\    
\cline{4-7}
&&&Modern &NOE &GOE &Prebiotic \\
\hline
$\sqrt[4]{a_4}$ &P-T Parameter (Degree 4)     & $\mathcal{U}(0.5,1.8)$ &0.694 & 1.117 & 0.999 & 1.189 \\ 
$a_3$           &P-T parameter (Degree 3)  &  $\mathcal{U}(0,100)$  & 18.644 & 17.562 & 12.886 &19.430\\
$a_2$           &P-T Parameter (Degree 2)  &  $\mathcal{U}(0,500)$   & 110.331 & 72.414 & 59.984 & 71.778\\
$a_1$           &P-T Parameter  (Degree 1) &  $\mathcal{U}(0,500)$   & 177.997 & 140.294 &123.764 &120.546\\
$a_0$           &P-T Parameter (Degree 0) &  $\mathcal{U}(0,1000)$  & 293.601 & 307.747 & 290.306 &280.615\\

$\log_{10}\left(P_0\left[\mathrm{bar}\right]\right)$&Surface Pressure    & $\mathcal{U}(-4,3)$& 0.007 & 0.017 & 0.008 & 0.008 \\
$R_{\text{pl}}\,\left[R_\oplus\right]$&Planet Radius (bulk value)&  $\mathcal{G}(1.0,0.2)$ & 1.000 & 1.000 & 1.000 &1.000\\ 
$\log_{10}\left(M_{\text{pl}}\,\left[M_\oplus\right]\right)$&Planet Mass (bulk value) & $\mathcal{G}(0.0,0.4)$& 0.000& 0.000 & 0.000 &0.000\\

$\log_{10}(\mathrm{N_2})$    &\ce{N2}  Mass Fraction  & $\mathcal{U}(-15,0)$  & -0.113 & -0.016 &  -0.005& -0.073 \\
$\log_{10}(\mathrm{O_2})$    & \ce{O2} Mass Fraction     & $\mathcal{U}(-15,0)$  & -0.631 & -1.623 & -2.622 & -6.002\\
$\log_{10}(\mathrm{H_2O})$   & \ce{H2O} Mass Fraction           & $\mathcal{U}(-15,0)$ & -2.607 & -2.143 & -2.498 & -2.345\\ 
$\log_{10}(\mathrm{CO_2})$   & \ce{CO2} Mass Fraction     & $\mathcal{U}(-15,0)$ & -3.265 & -1.807 & -1.806 &-0.827\\
$\log_{10}(\mathrm{CH_4})$   & \ce{CH4} Mass Fraction         & $\mathcal{U}(-15,0)$ & -6.028 & -3.635 & -3.035 & -6.056\\
$\log_{10}(\mathrm{O_3})$    & \ce{O3} Mass Fraction            & $\mathcal{U}(-15,0)$ & -6.436 & -6.567 & -7.412 & -10.301\\
$\log_{10}(\mathrm{CO})$     &\ce{CO}  Mass Fraction       & $\mathcal{U}(-15,0)$& -7.215 & -6.072 & -7.206 & -4.915\\
$\log_{10}(\mathrm{N_2O})$   &\ce{N2O} Mass Fraction     & $\mathcal{U}(-15,0)$& -6.343 &-6.859 & -7.962 &-\\

\hline 
\end{tabular}
\tablefoot{$\mathcal{U}(x,y)$ denotes a boxcar prior with a lower threshold $x$ and upper threshold $y$; $\mathcal{G}(\mu,\sigma)$ represents a Gaussian prior with mean $\mu$ and standard deviation $\sigma$. For $a_4$ we choose a prior on $\sqrt[4]{a_4}$, which allows us to sample small values more densely, typical of a fourth order coefficient, and then take the fourth power to obtain $a_4$. The expected values of the P-T parameters are generated by fitting the input P-T profiles \citep[calculated by][]{Rugheimer2018} with a fourth degree polynomial. The expected values of the abundances are calculated considering the weighted mean by pressure of the input abundance profiles \citep[calculated by][]{Rugheimer2018}.}
\end{table*}

\subsection{Assumptions and discrepancies}\label{sec:limitations}

When performing retrievals, we are limited by the maximum number of parameters retrieved in a reasonable computing time. For this reason, we need to make a few simplifications:

\begin{enumerate}
    \item As in Paper III, we parameterize the P-T profile in the retrieval by a fourth order polynomial.
    \item  We assume the abundances of the considered species to be independent of altitude. 
    \item  Our retrieval framework does not model clouds for any of the considered scenarios. This is a strong simplification for the cases where we retrieve cloudy input spectra. However, this retrieval approach allows us to investigate the biases on the results obtained when retrieving a cloudy spectrum assuming a cloud-free atmosphere. The addition of a cloud model to our retrieval framework will be tackled in a future study. 
    \item  A surface reflectance of 0.1 is assumed for all the wavelength range. This is a common value for water-rich habitable terrestrial planets, dominated by the low IR reflectance of oceans and ice.
\end{enumerate}

In contrast, \citet{Rugheimer2018} used P-T profiles that were self-consistently calculated by their climate-photochemistry model to generate the input spectra. The chemical abundance profiles were also altitude-dependent and calculated self-consistently by the climate-photochemistry model.
In order to compare the results of the retrievals with the input, we approximate the input P-T and abundance profiles to calculate expected values of the parameters considered in the retrievals.
 Regarding the thermal structure of the atmosphere, we determine the expected values of the polynomial coefficients ($a_4,\ a_3,\ a_2,\ a_1,\ a_0$) listed in Table~\ref{table:parameters} by fitting a fourth-order polynomial to the self-consistent P-T profiles. As for the abundances, we assume the weighted means over the pressure grid of the altitude-dependent abundance profiles. This is sensible, since the denser layers of the atmosphere (which corresponds to higher pressures) are generally the ones that contribute more to the spectrum.  The expected values for the considered species are listed in Table~\ref{table:parameters} as well.

For what concerns the surface reflectance, \citet{Rugheimer2018}  considered a wavelength-dependent reflectance. However, the reflectance still averaged to 0.1 in the wavelength range of interest. We would therefore not expect large variations due to this parameter.

There are also differences between the opacity tables that the two models use. We used the default set of opacities for \texttt{petitRADTRANS} as presented in \citet{2019A...627A..67M}. We added the \ce{N2O} opacity from  the ExoMol database \citep{2021A...646A..21C}. The CIA opacities are taken from the \emph{HITRAN} database. Details and reference papers corresponding to the opacity linelists are shown in Tables \ref{table:opacities} and \ref{table:cia}. In contrast, the input spectra were calculated using \emph{HITRAN 2016} opacities \citep{Rugheimer2018}. Differences in line-lists and broadening coefficients are therefore to be expected and may cause biases in the results. 

Due to these differences in the atmospheric models, we would imagine to find some small discrepancies between the spectra that were published in \citet{Rugheimer2018} and the ones that our framework can calculate. We will discuss this particular aspect in Section~\ref{sec:systematics}.

\begin{table*}[!htbp]

\caption{References for the molecular opacities used in the retrievals. }             
\label{table:opacities}      
\centering                          
\begin{tabular}{llll}    
\hline\hline                 
Species & Line List & Pressure Broadening & Line Cutoff\\
\hline

\ce{CO2}   & HITEMP \citep{ROTHMAN20102139} & $\gamma_{\mathrm{air}}$ &  \citet{Burch:69}  \\
 \ce{O3}  & HITRAN 2012 \citep{2013JQSRT.130....4R} & $\gamma_{\mathrm{air}}$ & \citet{2002JQSRT..72..117H}  \\
 \ce{CH4}   & ExoMol 2014  \citep{2014MNRAS.440.1649Y} & Eq. 15 \citep{2007ApJS..168..140S}  &  \citet{2002JQSRT..72..117H} \\
 \ce{CO}  & HITEMP \citep{ROTHMAN20102139}   &$\gamma_{\mathrm{air}}$ &  \citet{2002JQSRT..72..117H} \\
 \ce{H2O}   & HITEMP \citep{ROTHMAN20102139}  &$\gamma_{\mathrm{air}}$ & \citet{2002JQSRT..72..117H} \\
 \ce{N2O}  & ExoMol \citep{2021A...646A..21C}  & $\gamma_{\mathrm{H}},\ \gamma_{\mathrm{He}}$  & Eq. 6 \citep{2021A...646A..21C} \\
\hline 
\end{tabular}
\tablefoot{Adapted from \citet{2019A...627A..67M}.}
\end{table*}

\begin{table*}[!htbp]

\caption{References for the CIA and Rayleigh opacities used in the retrievals. }             
\label{table:cia}      
\centering                          
\begin{tabular}{ll|ll}    
\hline\hline                 
CIA  & Reference & Rayleigh & References\\
\hline

\ce{N2-N2}   & HITRAN \citep{KARMAN2019160} & \ce{N2} &  \citet{2014JQSRT.147..171T,2017JQSRT.189..281T}  \\
 \ce{O2-O2}  & HITRAN \citep{KARMAN2019160} & \ce{O2} & \citet{2014JQSRT.147..171T,2017JQSRT.189..281T}  \\
 \ce{O2-N2}   & HITRAN \citep{KARMAN2019160} & \ce{CO2}  &  \citet{2005JQSRT..92..293S} \\
 \ce{CO2-CO2}  & HITRAN \citep{KARMAN2019160}   & \ce{H2O} &  \citet{1998JPCRD..27..761H} \\
  &  & \ce{CH4} & \citet{2005JQSRT..92..293S} \\
   &  & \ce{CO} & \citet{2005JQSRT..92..293S} \\
\hline 
\end{tabular}
\tablefoot{Adapted from \citet{2019A...627A..67M}.}
\end{table*}

\section{Results}\label{sec:results}

In this section we show the results from the retrievals on the grid of different input spectra (see Table~\ref{table:id}) assuming the baseline parameters listed in Table~\ref{tab:lifesim}. We start by analyzing the retrieved spectra (Section~\ref{sec:spectra}) to offer a broad overview of the retrieval performance. Then, we study the retrieved P-T profiles (Section~\ref{sec:ptprofiles}), the planetary parameters (Section~\ref{sec:planetaryparameters}) and abundances (Section~\ref{sec:abundances}).
We also ran additional retrievals of the same scenarios (shown in Table~\ref{table:id}) by varying R and S/N. We compare the results of these retrievals in Section~\ref{sec:r-snr}.

\subsection{Retrieved emission spectra}\label{sec:spectra}
   
\begin{figure*}
   \centering
 \includegraphics[width=\textwidth]{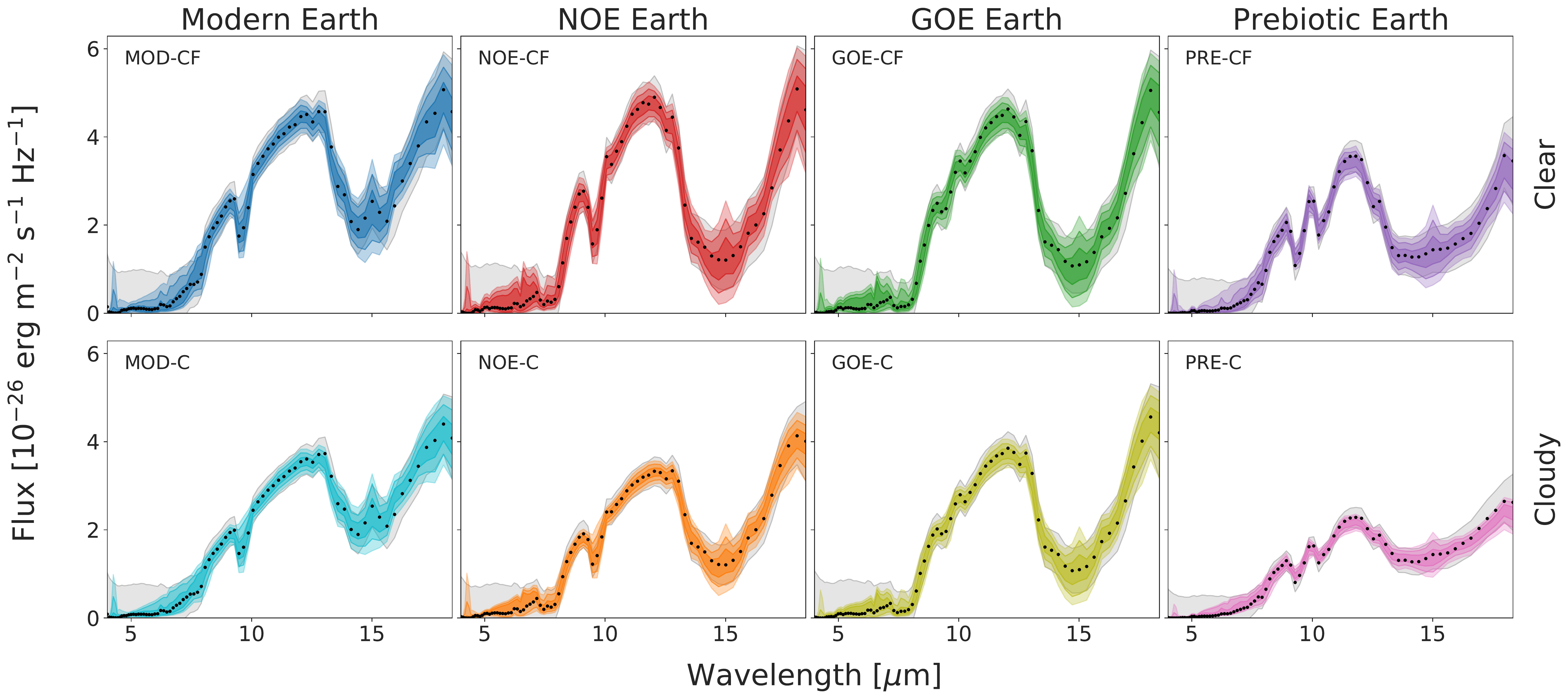}
   
   \caption{Retrieved spectra compared to the input spectra (black dots) for the various scenarios, ordered by epoch (columns) and cloud coverage (rows). The gray shaded area indicates the LIFE\textsc{sim} uncertainty. The color-shaded areas represent the confidence envelopes (darker shading corresponds to a higher confidence). The scenarios are color-coded according to Table~\ref{table:id}. }
    \label{fig:RetSpectra}
\end{figure*}

\begin{figure*}
   \centering
 \includegraphics[width=\textwidth]{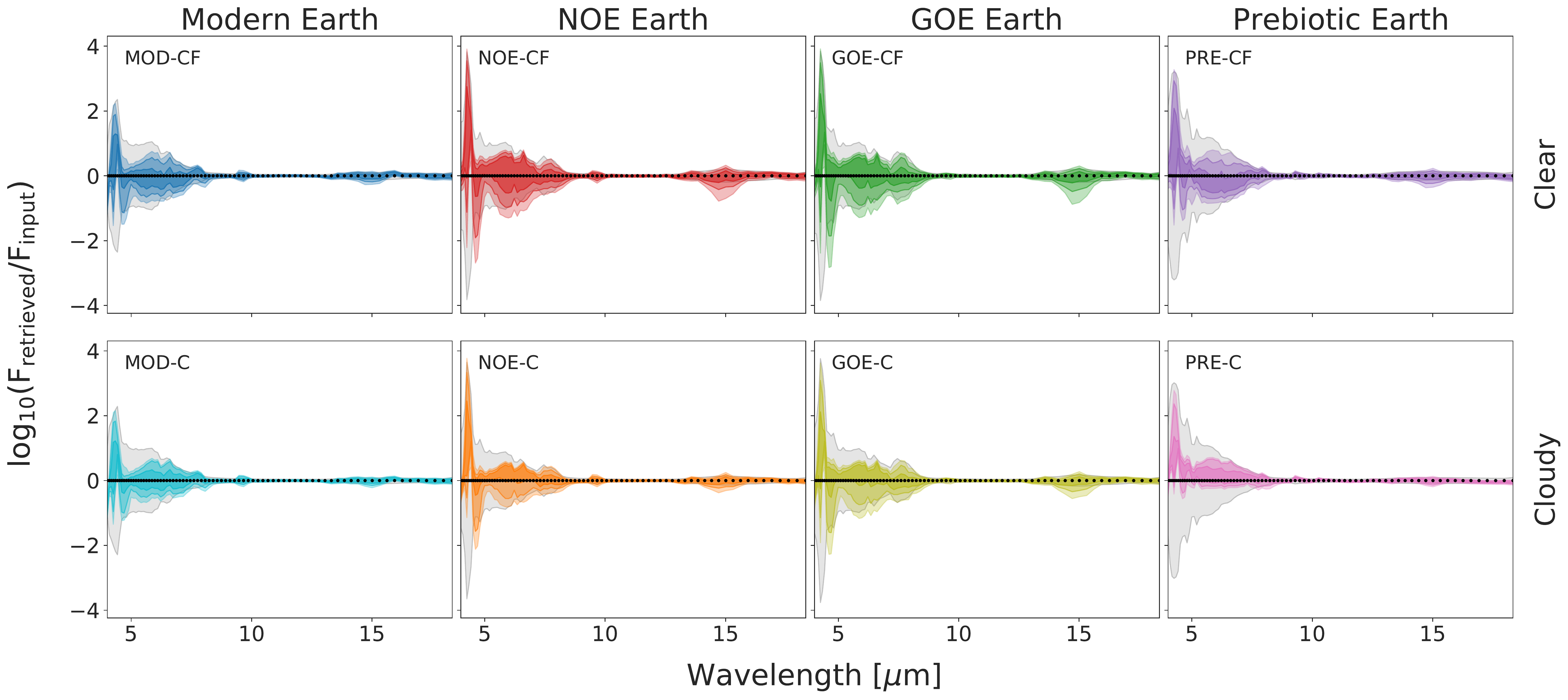}
   
   \caption{Ratios between the retrieved flux and the input flux (in logarithmic scale) for the various scenarios, ordered by epoch (columns) and cloud coverage (rows). The gray shaded area indicates the LIFE\textsc{sim} uncertainty. The color-shaded areas represent the confidence envelopes (darker shading corresponds to a higher confidence). The scenarios are color-coded according to Table~\ref{table:id}.  }
    \label{fig:RetSpectraResiduals}
\end{figure*}

The main output of the Bayesian retrieval framework are the posterior distributions of the parameters, necessary to produce theoretical spectra, that best match the data. 
The posteriors can be visualized as an $N$-dimensional space that is a subset of the larger $N$-dimensional prior space, $N$ being the number of parameters. Each point included in the posterior space has $N$ coordinates and represents a combination of $N$ parameters that, if fed to the theoretical spectral model, would produce a spectrum that was determined by the Bayesian framework to resemble the observed spectrum. 

From the available sets of parameters within the posteriors that the routine has calculated, we can therefore produce "retrieved spectra". These are shown in Figure~\ref{fig:RetSpectra}. Each subplot presents the results for a specific model. The input spectrum is binned down to R~=~50. Every flux point is represented by black dots and the uncertainty determined by LIFE\textsc{sim}  is shown as a gray shaded area. The retrieved spectra are color-coded according to Table~\ref{table:id}. The color of the shading is scaled according to the uncertainty in the retrieved spectra: the 1-$\sigma$ uncertainty region is shown in a darker color than the 2- and 3-$\sigma$ regions. Similarly, Figure~\ref{fig:RetSpectraResiduals} shows the logarithm of the ratio between the retrieved emission spectrum and the input emission spectrum for each scenario.

The retrieved spectra are generally in good agreement with the input spectra (within 1 $\sigma$) for all considered cases. This shows that our retrieval framework is able to reproduce the simulated input spectra, regardless of the complexity of the input model (in terms of thermal and abundance profiles, and cloud coverage). However, we notice regions with larger uncertainties, especially at wavelengths shorter than $\approx8\,\mu$m. Here, the ratio between the retrieved spectra and the input spectrum (as shown in Figure~\ref{fig:RetSpectraResiduals}) reaches up to a few orders of magnitude. Such differences are, however, still within the noise uncertainty (gray shaded areas).
Smaller differences can be noticed in the main \ce{CO2} band at $\approx15\,\mu$m. Since the noise is not as high as it is in the short wavelength range, these differences are probably due to discrepancies in the opacity tables (see Section~\ref{sec:systematics}).

The parameter estimation routine included in the Bayesian retrieval framework has the task of minimizing the difference between model output and data. For this, no detailed parameterization of the relevant physical processes are required.
Many of the relevant physical and chemical parameters are correlated (e.g. the planetary mass and the pressure, the pressure and the chemical abundances), often in a non linear way. It is therefore possible for the parameter estimation routine to produce similar spectra as the full (physical) input model over a diverse set of parameters, as a result of such correlations. It is appropriate therefore to question whether the parameter estimation routine results are on the one hand physically representative, or whether degeneracies and systematics between the retrieved parameters could be influencing the results. The next sections will explore these issues in more detail.

\subsection{Retrieved P-T profiles}\label{sec:ptprofiles}

\begin{figure*}
   \centering
    \includegraphics[width=\textwidth]{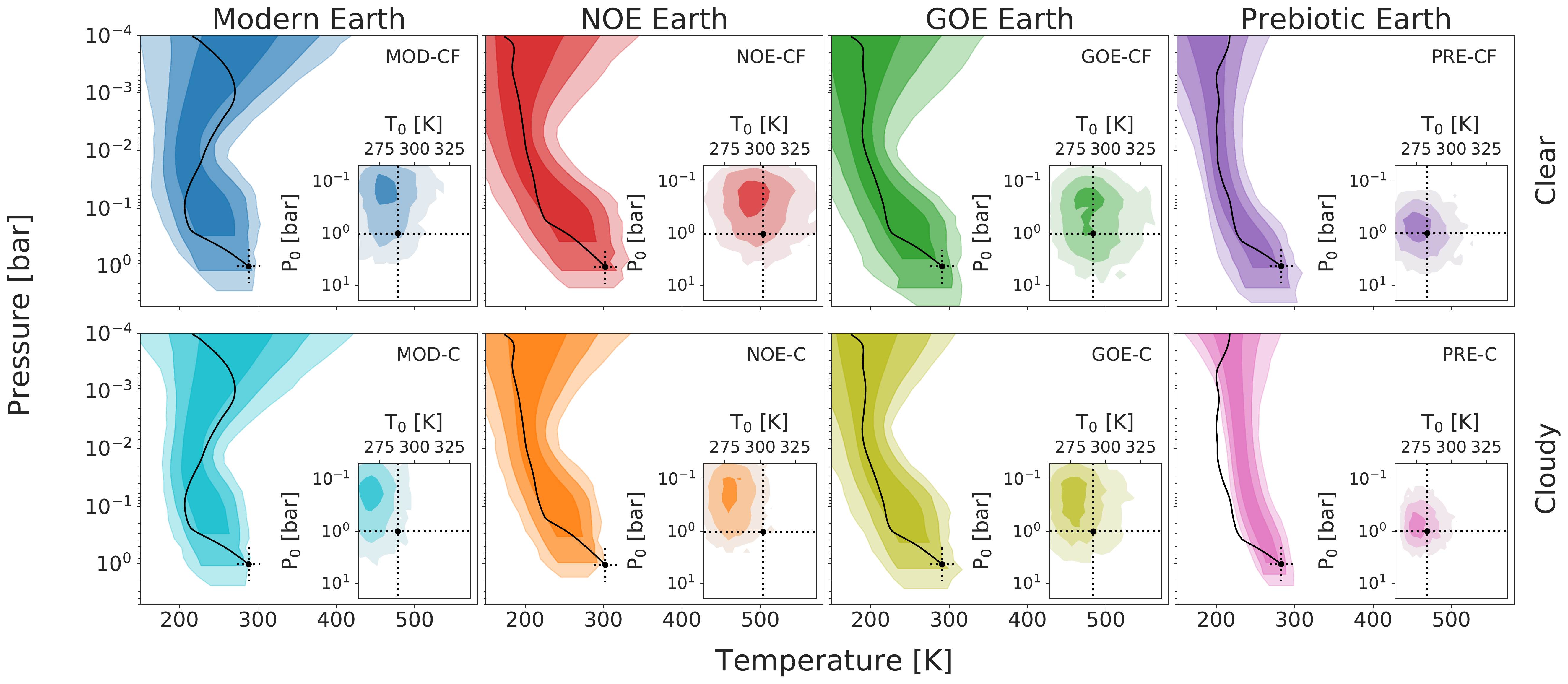}
   
   \caption{Retrieved P-T profiles compared to the input profiles (solid black line) for the various scenarios, ordered by epoch (columns) and cloud coverage (rows). The scenarios are color-coded according to Table~\ref{table:id}.  In each subplot, we also show an inset plot with the two-dimensional histogram of the retrieved surface P-T values. The 1-, 2-, and 3-$\sigma$ confidence levels in the P-T profiles and the 2D histogram are indicated by the increasing intensity of the color fill (darker shading correspond to a higher confidence). }
    \label{fig:RetProfiles}
\end{figure*}

In Figure~\ref{fig:RetProfiles} we show the retrieved P-T profiles compared to the input profiles for all combinations of the four epochs (columns) and the two cloud coverages (rows). 

The vertical shape of the retrieved P-T profiles in the lower atmosphere (pressures $\geq 10^{-2}$ bar) roughly follows that of the true P-T profiles. In most cases the true profiles are contained within the 1-$\sigma$ uncertainty envelope. As in Paper III, the uncertainties grow larger at higher altitudes (pressures $\leq 10^{-2}$ bar). This indicates that, for the quality of the input spectra we consider for this study, it is not possible to distinguish atmospheres with a stratospheric temperature inversion (i.e. the Modern Earth scenario) from those with an isothermal stratosphere (i.e. the NOE, GOE, and Prebiotic scenarios). This retrieval limitation is a result of the small overall contribution of the upper atmospheric layers to the planet's MIR emission spectrum.

A few additional inconsistencies between the retrieved and input P-T profiles are also apparent in the lower altitudes (high pressures). A general feature in the three biotic epochs (Modern, NOE, and GOE Earth) is the retrieval of underestimated values for the ground pressure $P_0$ ($\sim$0.1 bar as opposed to the true value of $\sim$1 bar). This occurs for both the cloud-free and cloudy spectra. Such offset could be explained by systematic differences between the radiative transfer models used to produce and to retrieve the simulated spectra. We will discuss this in more detail in Section~\ref{sec:systematics}.
The ground temperatures $T_0$ are on average well retrieved for all clear sky scenarios. In contrast, the retrievals performed for the cloudy spectra systematically underestimate $T_0$, with differences between the retrieved and true value $\lesssim$ 25 K. These results have an impact on assessing the habitability of the simulated exoplanets, which will be discussed in more detail in Section~\ref{sec:discussion}.

For the Prebiotic Earth input spectra, the retrievals provide estimates for $P_0$ and $T_0$ that are in agreement with the true parameter values. Furthermore, the overall uncertainties on the retrieved P-T profile are generally smaller than for the other epochs. However, for the cloudy Prebiotic Earth (PRE-C) spectrum, the retrieved P-T profile is a few tens of Kelvin warmer than the true value in the intermediate layers of the atmosphere ($\sim10^{-1}$ to $\sim10^{-3}$ bar). This effect is likely related to the much weaker emission features in the cloudy Prebiotic Earth scenario compared to the other epochs. We will discuss the impact of neglecting clouds in the retrievals in Section~\ref{sec:clouds}.

\subsection{Retrieved planetary parameters}\label{sec:planetaryparameters}

\begin{figure*}
   \centering
   \includegraphics[width=0.95\linewidth]{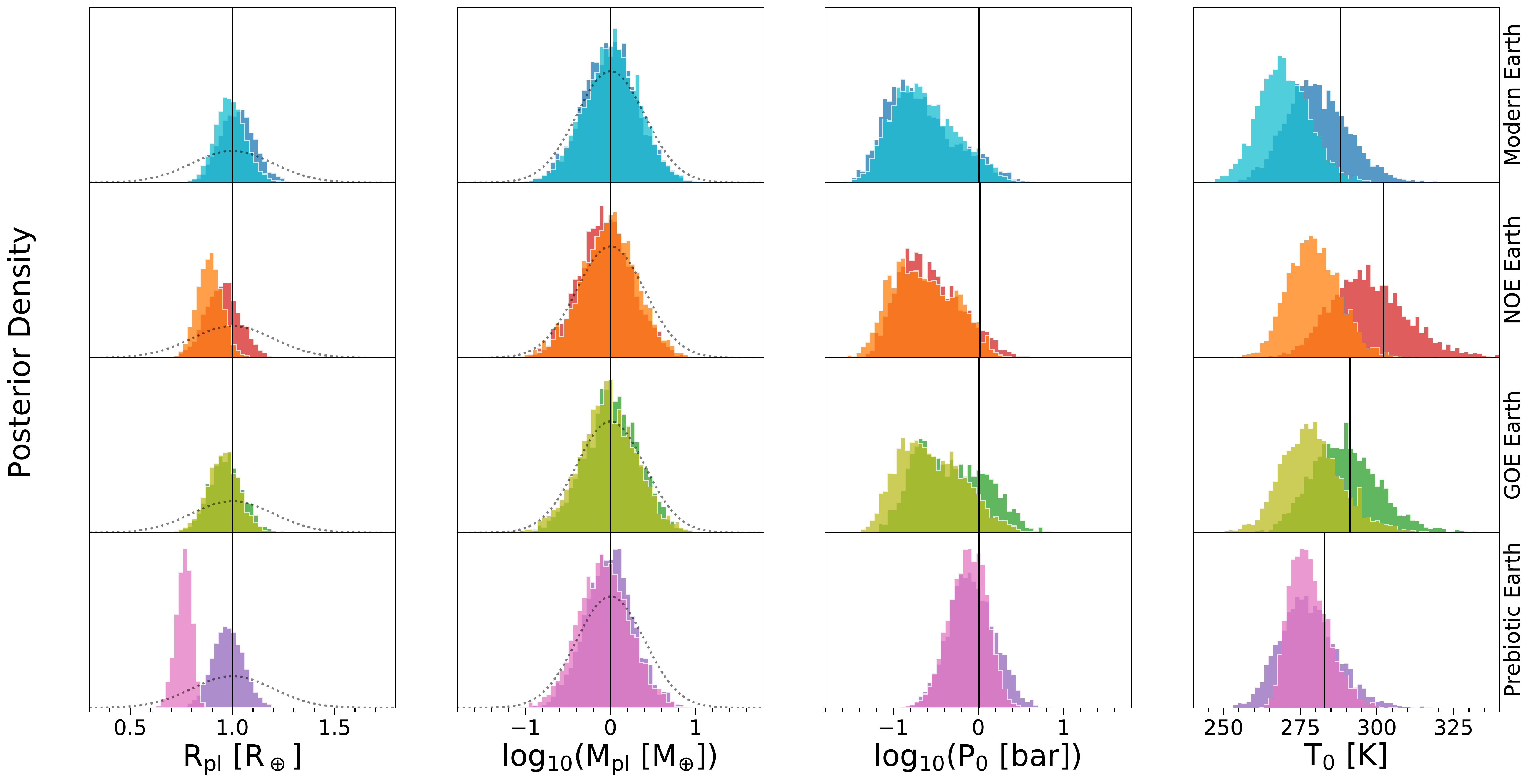}
      \caption{Posterior density distributions for the retrieved exoplanet parameters (columns) for the different epochs (rows) and cloud coverages. We follow the color-coding listed in Table~\ref{table:id} to differentiate the different scenarios.  
      The vertical, solid lines mark the true values for each parameter.
      The dotted lines in the $R_{\text{pl}}$ and $M_{\text{pl}}$ plots indicate the assumed Gaussian priors. For $P_0$ and $T_0$ we assume broad, flat priors which are not plotted.}
         \label{fig:parameters_epochs}
\end{figure*}
   
Figure~\ref{fig:parameters_epochs} shows the posterior density distributions we retrieve for the planetary parameters ($R_\mathrm{pl}$, $M_\mathrm{pl}$) and surface conditions ($P_0$, $T_0$) for all considered input spectra. $R_\mathrm{pl}$, $M_\mathrm{pl}$, and $P_0$ are directly retrieved by our framework, while the $T_0$ posterior is calculated from the P-T parameters (polynomial coefficients $a_i$) and $P_0$ posteriors. The results are color-coded based on Table~\ref{table:id} and grouped by epoch (rows) and planetary parameters (column). The retrieved parameters for the clear and cloudy input spectrum of the same epoch are shown in the same subplot to facilitate comparison. 

We obtain good estimates for all of the parameters considered, especially for the cloud-free scenarios. The posterior distributions roughly follow a Gaussian distribution and are typically centered on the true values. 

By comparing the retrieved posteriors for $R_\mathrm{pl}$ to the corresponding Gaussian prior range (shown in Figure~\ref{fig:parameters_epochs} as a dotted lines), we observe that we manage to better constrain the planet radius $R_\mathrm{pl}$ with respect to the prior distribution. For the GOE and Modern Earth scenario, the retrieved posteriors do not significantly depend on the cloud coverage. For the NOE and Prebiotic input we underestimate $R_\mathrm{pl}$ in the retrievals of the cloudy spectra (retrieved as $\sim$ $0.8-0.9$ $R_\oplus$ instead of 1 $R_\oplus$). This difference is more pronounced in the Prebiotic scenario. 

For what concerns $M_\mathrm{pl}$, the retrieval analysis does not add further constraints on the estimates for the planet's mass. Also, there is no noticeable difference in the retrieval results for $M_\mathrm{pl}$ between the four epochs. This finding holds for both the clear and cloudy scenarios and is in agreement with the results we presented in Paper III. 
It is important to note that, in contrast to the prior assumption (see Section \ref{sec:framework}), $M_\mathrm{pl}$ is not linked to $R_\mathrm{pl}$ through a mass-radius relationship during the retrieval, but it is instead a free parameter. This means that both $M_\mathrm{pl}$ and $R_\mathrm{pl}$ are independently drawn from their respective prior. In the retrieval, we use both parameters to calculate the planet's surface gravity, which is required to compute the theoretical emission spectrum. In addition to the surface gravity calculation, we also use $R_\mathrm{pl}$ to scale the flux emitted per unit area at the top of the atmosphere (as calculated by \texttt{petitRADTRANS}) to the observed exoplanet flux at a distance $d$ from the observer (generally well known, 10 pc in our study). We do so by multiplying the flux at the top of the atmosphere by the factor $({R_\mathrm{pl}}/d)^2$.

Since the surface gravity is typically not directly constrainable due to the gravity-abundance degeneracy (see Section \ref{sec:r-snr}), the retrieval struggles to further constrain $M_\mathrm{pl}$. In contrast, the retrieval does manage to constrain $R_\mathrm{pl}$ further, as it does not only dependent on the surface gravity, but also on the distance-scaling of the spectrum.

The retrieved posteriors for the surface pressure $P_0$ are significantly smaller than the assumed prior distribution ($10^{-4}$ to $10^3$ bar), meaning that we manage to pose strong constraints on $P_0$ with respect to the assumed prior knowledge on the parameter. However, the retrieval tends to underestimate the value of $P_0$ in all cases except for the Prebiotic one. The retrieved posteriors in these cases are not well represented by a Gaussian, which indicates that the retrieval results for $P_0$ could be degenerate. We will discuss this in more detail in Section~\ref{sec:quality}.

As shown in Section~\ref{sec:ptprofiles}, the retrieved posteriors for the surface temperature $T_0$ are centered on the true values for the cloud-free retrievals. For the cloudy input spectra, the retrievals tend to underestimate the surface temperature.  The standard deviation of the retrieved $T_0$ posteriors is roughly $\pm20$ K for all biotic scenarios. This is in agreement with the findings made in Paper III.  
The spread of retrieved posteriors could potentially be reduced by increasing the R or S/N of the input spectra (see Section \ref{sec:r-snr}). Observing strategies for the trade-off between R and S/N will be addressed in Section~\ref{sec:quality}.

\subsection{Retrieved chemical abundance parameters} \label{sec:abundances}

  \begin{figure*}
   \centering
   \includegraphics[width=0.95\linewidth]{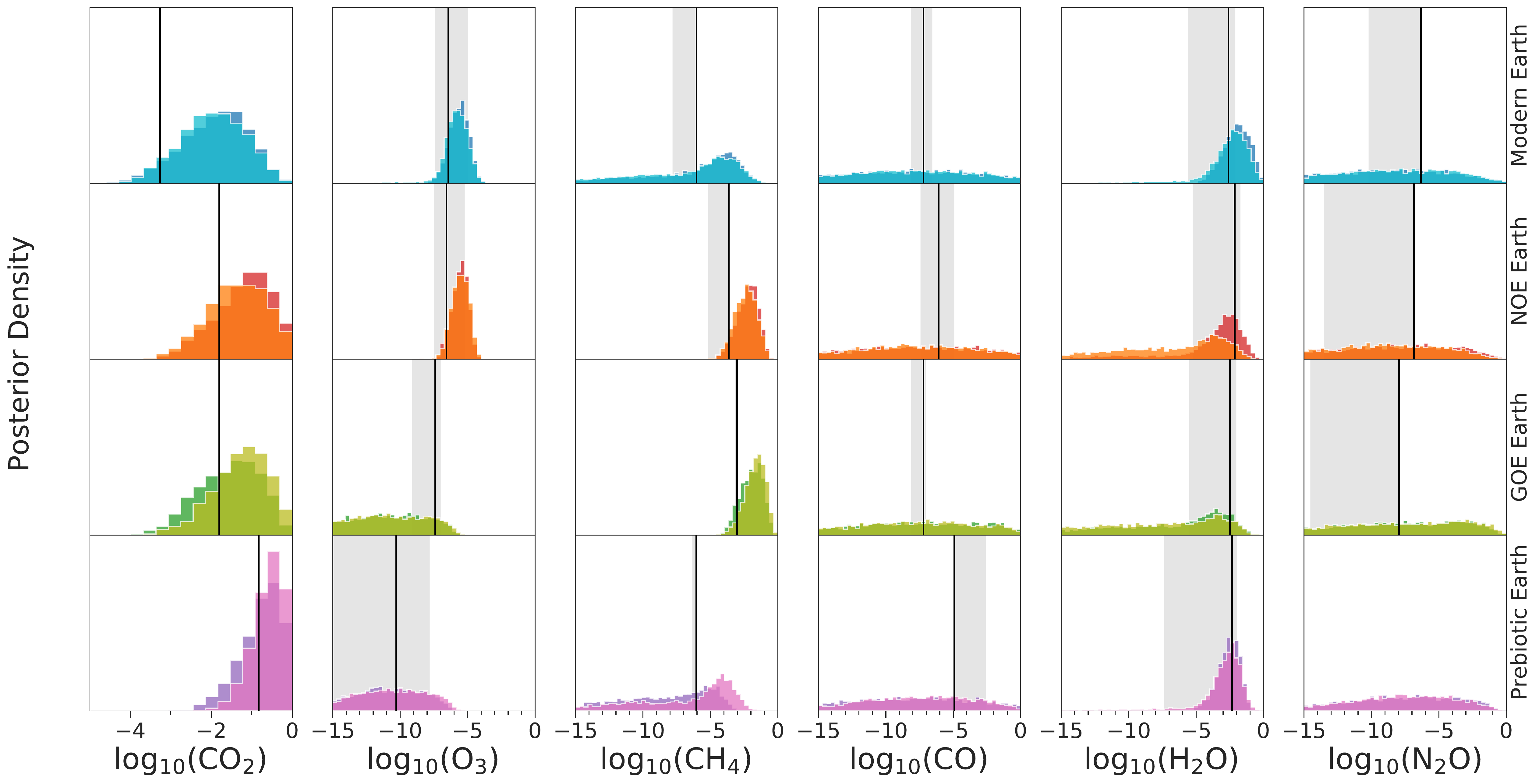}
      \caption{
      Posterior density distributions for the retrieved species (columns), for the different epochs (rows), and cloud coverage scenarios.
      Results from the various scenarios use the color-coding from Table~\ref{table:id}. The solid black lines indicate the expected values for each species, which vary depending on the epoch. The gray shaded area marks the range of values in the vertically non-constant abundance profiles which feature in the input spectra.}
         \label{fig:abundances_epochs}
   \end{figure*}

Figure~\ref{fig:abundances_epochs} shows the retrieved posterior distributions for the main atmospheric gases. We again arrange the various scenarios by epoch (row), and atmospheric species (column) and use the color-coding from Table~\ref{table:id}. The results for the clear and the cloudy retrievals of one epoch are shown in the same subplot to facilitate comparison. 

We plot our expected abundances (listed in Table \ref{table:parameters}), which are the weighted means (with respect to the pressure) of the original abundance profiles, as black vertical lines. If no true value is plotted, the molecule is not present in the input spectrum. We further indicate the range of variability of the true, pressure-dependant abundance profiles (minimum to maximum) via the shaded gray area in each subplot. 

We adopt the same posterior classification scheme that was introduced in Paper III, for an easier comparison of the results. This scheme divides the retrieved posteriors into the following four classes:

\begin{itemize}
    \item Constrained (C): The posterior is best described by a Gaussian distribution. This implies that abundances both significantly lower and higher than the true value can be ruled out.
    
    \item Sensitivity limit (SL): The abundance is at the retrieval's detection limit for the species. The posterior exhibits a distinct peak. However, low abundances are not ruled out. The posterior is best described by the convolution of a soft-step function with a Gaussian.
    
    \item Upper Limit (UL): The posterior resembles a soft-step function. Large abundances can be excluded, low ones cannot.
    
    \item Unconstrained (UC): We cannot retrieve information on the atmospheric abundance. The posterior resembles a constant function over the full prior range.
\end{itemize}

For further details on the specifics of the posterior classification we refer the reader to the Appendix~B of Paper III.

We obtain UC posteriors for the abundances  of \ce{N2} and \ce{O2} in all retrievals performed. In accordance with the findings presented in Paper III, these molecules are not detectable in any of the considered scenarios. This finding indicates that the corresponding CIA spectral signatures are too weak to be detectable in the considered input spectra with R~$= 50$ and S/N~$= 10$. To increase readability, we choose not shown the retrieval results for \ce{N2} and \ce{O2} in Figure~\ref{fig:abundances_epochs}. The posterior distributions of these molecules can however be found in the corner plots in Appendix~\ref{app:corner}. 
Similarly, the trace gases \ce{N2O} and \ce{CO} were not detected in any of our retrievals, obtaining unconstrained posteriors for all epochs. The MIR absorption features of these molecules at the considered abundances are also too weak in the considered input spectra to be constrained in our retrievals. This can be seen by the flat posterior distributions for both species in all considered cases (see Figure~\ref{fig:abundances_epochs}).

We detect \ce{CO2} in all retrievals and the received posterior distributions are generally Gaussian-like (C-type posteriors). Our results suggest that the median abundances of the different posterior distributions are higher than the true value for all the epochs. However, in the Prebiotic, GOE, and NOE scenarios the true abundances still lie within the 1-$\sigma$ envelope of the retrieved abundances. For the Modern Earth scenarios, the true value lies within the 3-$\sigma$ range of the retrieved posterior. This is consistent with a "compensation effect" whereby the retrieval framework is correcting for the underestimated pressure. The degeneracy between chemical composition and atmospheric pressure is well known and it was already encountered in Paper III (see Section~\ref{sec:CompIII}).
All retrieved \ce{CO2} posteriors span about 3 orders of magnitude (3 dex). They all appear very similar even though the expected values of \ce{CO2} span from roughly 0.01\% (Modern Earth) to the order of 10\% (Prebiotic Earth). This forbids the use of \ce{CO2}, one of the major absorbers in the atmosphere, as discriminator between the considered epochs. To reduce the variance in the retrieved abundances, an increase in R and/or S/N might be recommended (see Section~\ref{sec:r-snr}). 

For the remaining species (\ce{O3}, \ce{CH4}, and \ce{H2O}) the retrieval results depend on the considered epoch: 

\ce{O3} is retrieved accurately in both the Modern Earth and NOE Earth scenarios (C-type posteriors). These are the two cases where \ce{O3} is more abundant ($\sim10^{-6}$ in mass fraction). In contrast, for the Prebiotic and GOE scenario, we only retrieve upper limits (UL-type posteriors) for the \ce{O3} abundance. This means that we can rule out high abundances of \ce{O3} ($\gtrsim10^{-6}$ in mass fraction), but cannot exclude abundances below the retrieved upper limit. 

We manage to detect \ce{CH4} in the NOE and GOE Earth spectra (C-type posterior), which have a higher \ce{CH4} abundance. For the Modern and Prebiotic Earth spectra, where the abundances are lower, we retrieve an SL-type posterior, which is characterized by a peak in the distribution roughly at the true value and a non-negligible tail towards low abundances. Similar to \ce{CO2}, \ce{CH4} is generally overestimated when detected. However, the true value still lies within the 2-$\sigma$ envelope the retrieved posterior distribution. The conditional retrieval of \ce{O3} and \ce{CH4} is particularly significant when discussing the detectability of biosignatures in Earth-like planets with \emph{LIFE}, which will be discussed in Section \ref{sec:differentiating}.

\ce{H2O} is constrained in both the Modern and Prebiotic scenarios, as well as the clear NOE Earth model, and the retrieved posteriors are centered on the true value. In contrast, we only detect SL-type posteriors for both the GOE Earth spectra and the cloudy NOE model. In this case, such a difference in retrieval performance cannot be explained by a difference in the abundance of the species, which is fairly constant throughout all the epochs (around $\sim10^{-2}$ in mass fraction). This is likely to be related to the overestimation of \ce{CH4}, a species that is much more abundant in the GOE and NOE scenarios ($10^{-3}$ in mass fraction compared to $10^{-6}$ for the Modern and Prebiotic epochs). The spectral signature of \ce{H2O} overlaps with that of \ce{CH4} between 5-7 $\mu$m, which is where the noise level is very high and the flux levels are low. Therefore, the retrieval framework favors a higher abundance of \ce{CH4} at the expense of a larger uncertainty on \ce{H2O} in these scenarios.

Finally, our results suggest only modest differences in the retrieved abundances obtained from the clear and cloudy input spectra. The presence of clouds in the spectrum does not seem to deteriorate the abundance estimation capabilities for most molecules. In contrast, as seen in the previous subsection, characterization of a cloudy atmosphere with a cloud-free model will likely result in biases in the retrieved physical parameters. We will discuss this topic further in Section~\ref{sec:clouds}.

\subsection{Runs at higher resolution and/or signal-to-noise ratio}\label{sec:r-snr}

In this Section, we investigate whether our retrieval results can be improved by increasing the quality (R and/or S/N) and thus the information content of the input spectra. We ran ancillary retrievals for the 8 scenarios and choose the following combinations of R and S/N\footnote{We remind the reader that the S/N refers to the value at a reference wavelength of 11.2~$\mu$m.}:
\begin{itemize}
    \item R~$=50$ and S/N~$=10$ (the reference case); 
     \item R~$=100$ and S/N~$=10$;
     \item R~$=50$ and S/N~$=20$;
    \item R~$=100$ and S/N~$=20$;
\end{itemize}

\begin{figure}
   \centering
    \includegraphics[width=\linewidth]{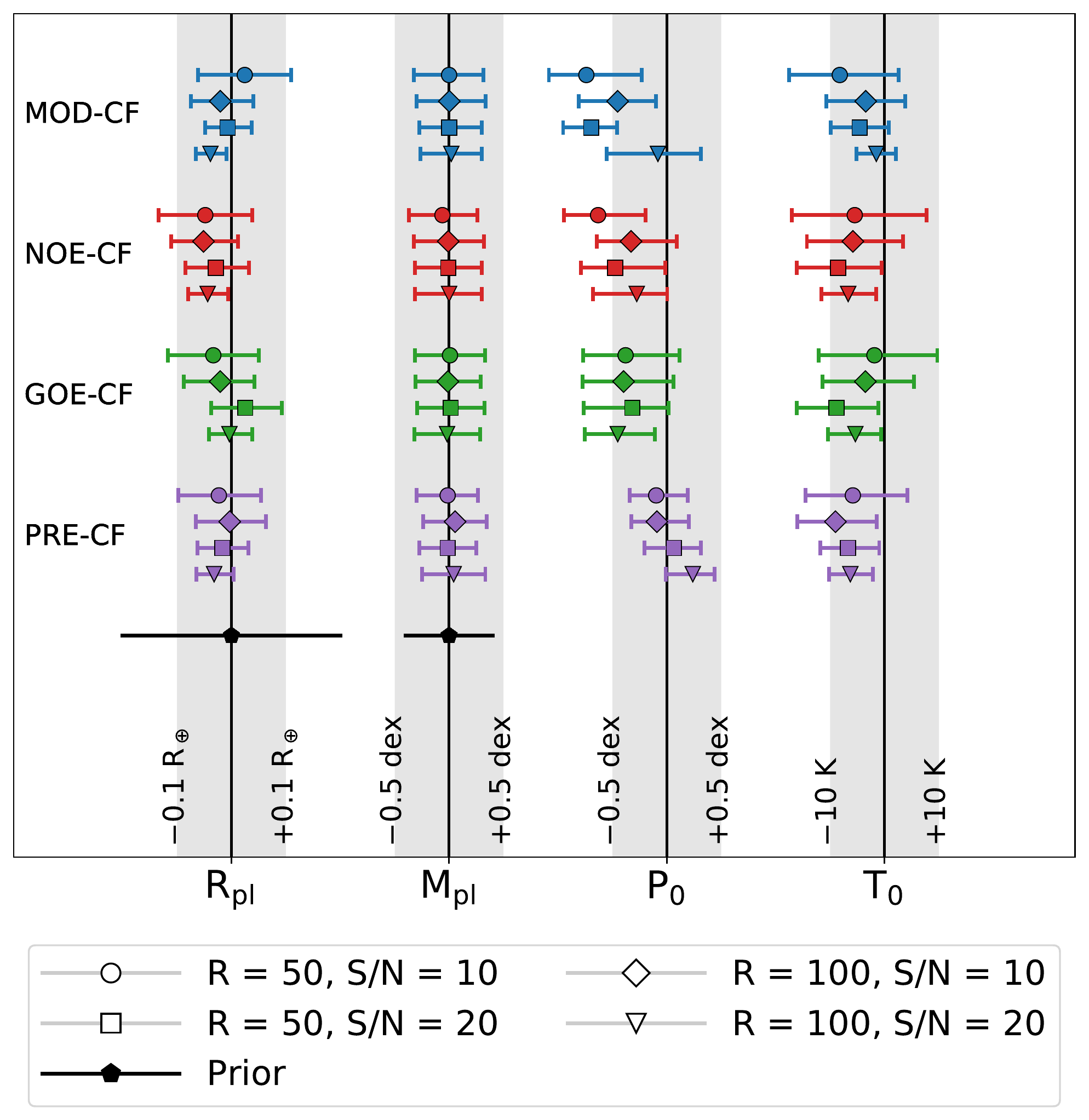}
   
   \caption{Retrieved exoplanet parameters for the different scenarios with varying R and S/N values. The error bars denote the 68\% confidence intervals. For $M_\mathrm{pl}$ and $R_\mathrm{pl}$, we also plot the assumed prior distributions. For $T_0$ and $P_0$, we assumed flat, broad
priors. The vertical lines mark the true parameter values.}
    \label{fig:QualityParams}
\end{figure}

\begin{figure*}%
    \centering
    \subfloat{{\includegraphics[width=.45\linewidth]{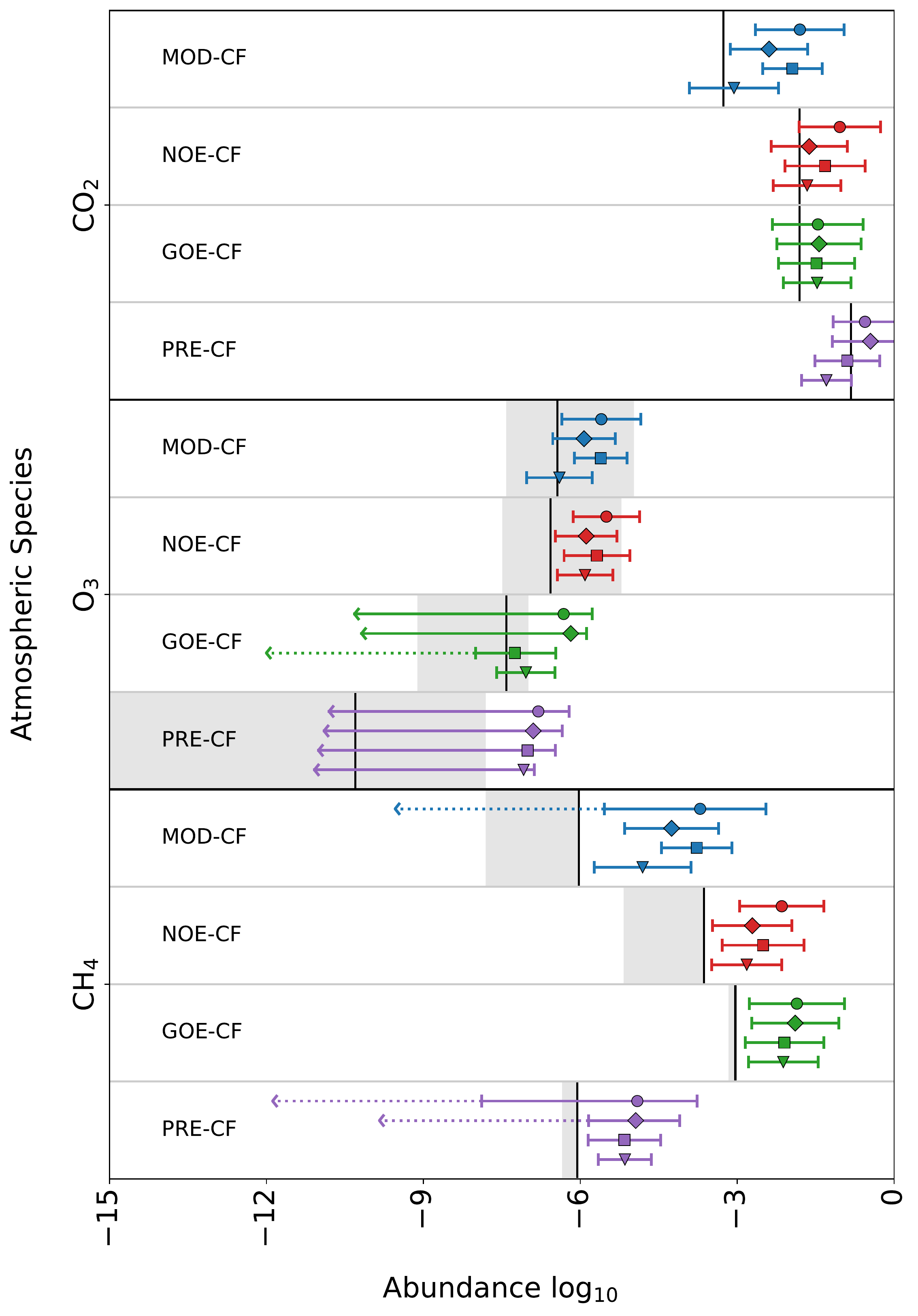} }}%
    \qquad
    \subfloat{{\includegraphics[width=.45\linewidth]{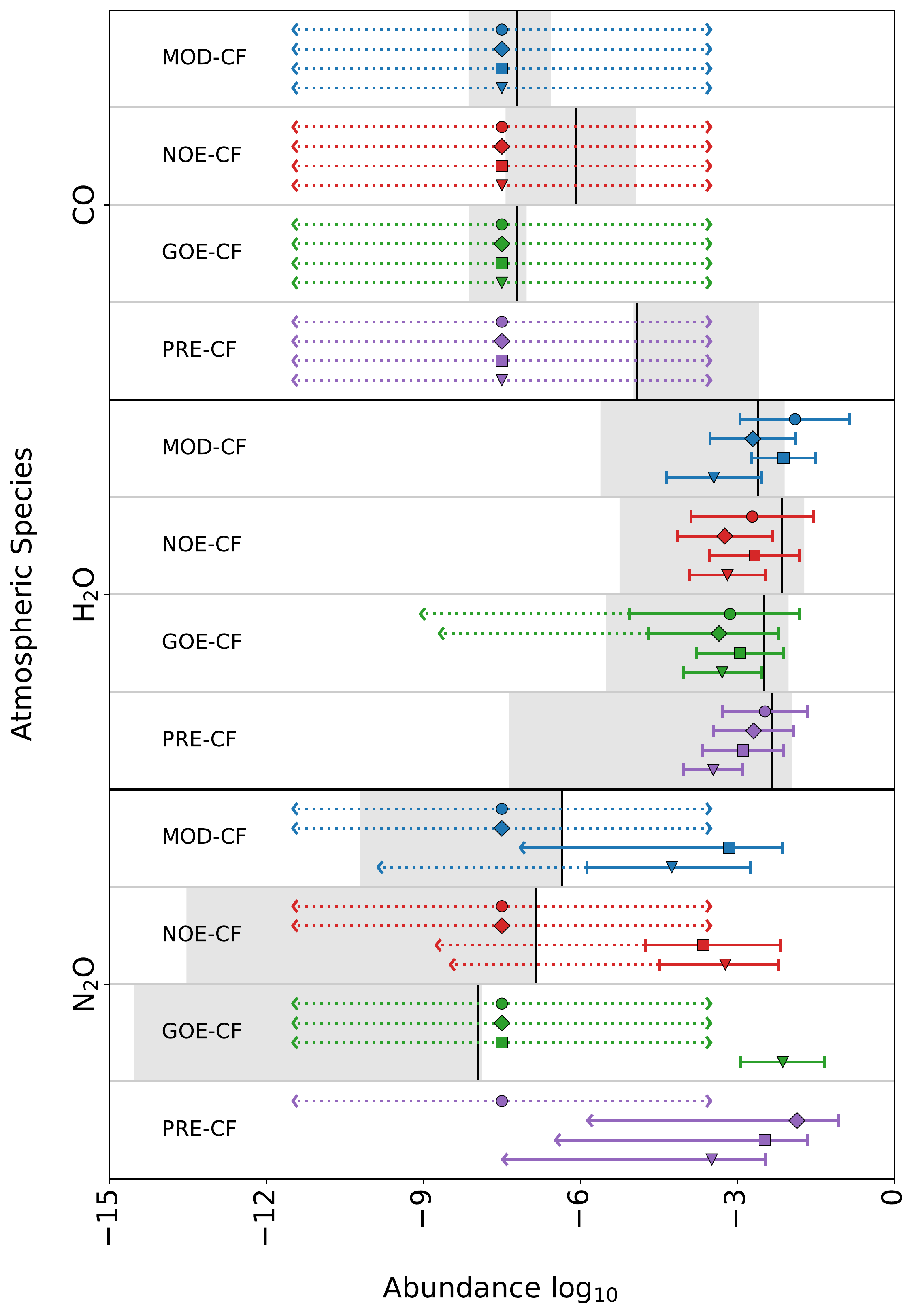} }}%
    \,
    \subfloat{\includegraphics[width=.4\textwidth]{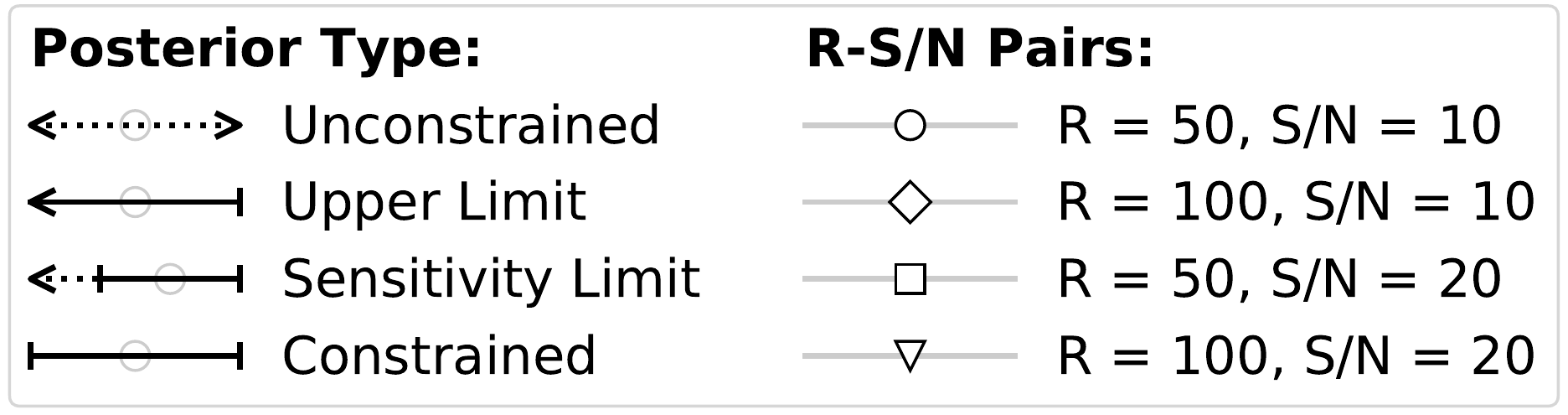}}
    \caption{Retrieved atmospheric abundances for the different ancillary runs. 
      Results belonging to the various scenarios are provided using the color-coding from Table~\ref{table:id}. We use different markers for the runs at different R-S/N (see legend). The solid lines indicate the expected values for each species, which vary depending on the epoch. The gray shaded areas mark the range of values in the vertically non-constant abundance profiles of the input spectra. The posterior distributions were classified using our posterior classification scheme (see Section~\ref{sec:abundances} for details).}
    \label{fig:QualityAbunds}%
\end{figure*}

We provide a summary of the results obtained from these additional retrieval runs in Figures \ref{fig:QualityParams} (planetary parameters) and \ref{fig:QualityAbunds} (abundances). Results from the different ancillary runs are represented using different markers. The results for the different epochs are color-coded according to Table~\ref{table:id}. Here, we only show the results for the clear input spectra. The plots corresponding to the cloudy scenarios can be found in Appendix~\ref{app:cloudy}.

We are particularly interested in significant increases in accuracy (i.e. the retrieved values agree better with the input "truth") or in precision (i.e. the posterior's variance is reduced).  For higher S/N we expect more precise results, since the uncertainty in the input spectrum is lower. This yields stronger constraints on the model parameters. An increase in R should allow for a more robust identification and characterization of the spectral features and thus more accurate retrieval results. Both increased accuracy and/or precision could allow us to differentiate between the different epochs (and generally between different planets) more clearly. This will be of great importance, especially when searching for signatures of life in exoplanetary atmospheres (see Section~\ref{sec:differentiating}).
We should point out that in our current simulation setup, which ignores (systematic) instrumental noise terms, doubling R at a constant S/N means doubling the integration time, while doubling S/N at a constant R means, roughly, quadrupling it\footnote{We refer the reader to the Appendix of Paper I, where we show a breakdown of the typical noise contributions for planets detected around Solar-type stars.}. This information is crucial for the mission planning and will be further discussed in Section~\ref{sec:quality}.

In Figure~\ref{fig:QualityParams}, we notice that increasing the S/N to $20$ while keeping ${\mathrm{R}=50}$ (square markers) generally results in a narrower posterior for $R_\mathrm{pl}$. We observe a reduction in variance of the $R_\mathrm{pl}$ posterior by up to a factor of 2 compared to the reference case (${\mathrm{R}=50}$, ${\mathrm{S/N}=10}$; the circular markers). An increase in R (${\mathrm{R}=100}$, ${\mathrm{S/N}=10}$; the diamond markers) causes the variance of the $R_\mathrm{pl}$ posterior to shrink to about 70\% of the reference case variance.   In contrast, we observe no noticeable gain in the accuracy of the retrieved value for $R_\mathrm{pl}$ when increasing S/N and R at the same time. On the other hand, the precision of the measurement at ${\mathrm{R}=100}$, ${\mathrm{S/N}=20}$ improves significantly, with the variance of the $R_\mathrm{pl}$ posterior shrinking up to three times compared to the reference case.

We further find that the retrieval of the planetary mass $M_\mathrm{pl}$ does not improve significantly when moving to higher R and S/N input spectra. We observe no significant increase in both accuracy and precision. This finding is consistent with the results shown in Paper III.  The underlying reason for this observation is the degeneracy between the surface gravity (and thus also $M_{\mathrm{pl}}$) and the abundances of trace gases \citep[see e.g. Paper III,][]{Molliere:Gravity_Abundance_Degeneracy,2018AJ....155..200F,Madhusudhan:Atmospheric_Retrieval,quanz2021atmospheric}.  Since gravity and abundances are involved in the hydrostatic equilibrium, it is possible to reproduce the same spectral feature using different combinations of these parameters. This broadens the variance of the posteriors of $M_\mathrm{pl}$ and of the atmospheric species.

Increasing the quality of the input spectrum does improve the accuracy of the retrieval for $P_0$ in the clear Modern Earth (MOD-CF) case. The results for the other epochs do not exhibit a similar trend with increasing input quality. This failure to retrieve accurate ground pressure estimates is likely rooted in differences between the opacity tables used by the retrieval framework and the ones used to calculate the input spectra (see Section~\ref{sec:systematics} for more details). Additionally, no noticeable decrease in the variance of the retrieved $P_0$ estimate is present for higher values of R or S/N. This is likely a result of the pressure-abundance degeneracy, which has already been described in Section~\ref{sec:abundances}.

For the surface temperature $T_0$, we do not notice any substantial improvements in the accuracy of the retrieved values when increasing R or S/N. However, as for $R_\mathrm{pl}$, we observe a significant reduction in the variance of the posteriors when increasing S/N and R. Compared to the reference case, the uncertainty in $T_0$ is reduced by a factor of 2 for the runs with S/N~$=20$ and to about 70\% of the reference variance for the runs with R~$=100$. These improvements in temperature accuracy could be crucial when assessing the potential habitability of an observed exoplanet.

In Figure~\ref{fig:QualityAbunds} we summarize the retrieved posterior distributions in the abundances for the reference case (R~$=50$ and S/N~$=10$, circular markers) and all other R and S/N combinations. The abundance posteriors are classified according to our classification scheme (see Section~\ref{sec:abundances} and Paper III).

Generally, we observe that increases in both S/N and R do not significantly improve the accuracy nor the precision of the retrieved posteriors for the majority of the scenarios. This is again the result of the pressure- and gravity-abundance degeneracies. In particular, the pressure-abundance degeneracy is responsible for the shifts with respect to the true values, whereas the gravity-abundance degeneracy defines the variance of the abundance posteriors. The effects of the pressure-abundance degeneracy can be noticed for \ce{CO2} in the clear Modern Earth (MOD-CF) scenario. For the reference case (circular marker) we strongly underestimated $P_0$, which is compensated by an overestimation in the \ce{CO2} abundance. As we move to higher R and S/N input spectra, our estimate for $P_0$ improves, which results in better accuracies for the retrieved \ce{CO2} abundance. The same connection between $P_0$ and the retrieved abundances can be seen for all other constrained species. 

In contrast, the variance of the \ce{CO2} posterior does not decrease significantly with increases in R and S/N since it is limited by the variance of the $M_\mathrm{pl}$ posterior (due to the gravity-abundance degeneracy), which is the same for all considered cases. While this behaviour describes the results for most species well, there are some noteworthy exceptions that we will discuss here.

Firstly, there could be a tentative detection of \ce{O3}  (an SL posterior) in the clear GOE Earth (GOE-CF) epoch when increasing the S/N to 20 (square marker). If also the resolution is increased to ${\mathrm{R}=100}$ (triangular marker), we could better constrain the \ce{O3} abundance. Purely increasing R to 100 would not improve the accuracy or the precision of \ce{O3} (diamond marker).
Similarly, increasing the S/N would allow for a detection of \ce{CH4} in all four epochs, which was not possible for the reference case (circular marker). However, the retrieved \ce{CH4} abundances are one to two orders of magnitude higher than the truths. Results suggest similar, but less pronounced systematic offsets with respect to the true values for the other constrained species. These offsets are likely the result of a combination of the degeneracy between $P_0$ and the abundances and systematic errors, such as differences in the molecular line lists (see Section~\ref{sec:systematics}).
Both \ce{O3} and \ce{CH4} are of particular interest for astrobiology, since they are indicative of disequilibrium chemistry in the atmosphere and could indicate the presence of biological activity on the planet. We will discuss this in more detail in Sections \ref{sec:quality} and \ref{sec:differentiating}.

Furthermore, an increase in S/N would enable a robust detection of \ce{H2O} in the clear GOE Earth (GOE-CF) epoch (C- instead of SL-type posterior). On the contrary, increasing the resolution alone does not have the same effect.

Finally, \ce{CO} is unconstrained for all epochs and R-S/N pairs, which indicates that this species could not be detected in an Earth-like atmosphere with \textit{LIFE}. Similarly, none of the runs were able to fully constrain the \ce{N2O} abundance. The retrieval can only provide upper limits on the \ce{N2O} abundance, which often only manage to rule out atmospheric abundances greater than 1\% in mass fraction. The retrieval is therefore not sensitive to these molecules -- their spectral signatures are too small compared to the LIFE\textsc{sim} noise to be detected even in the best considered scenario. An exception would be the clear GOE Earth at R~$=100$ and S/N~$=20$ scenario, for which we retrieve a wrong estimate of \ce{N2O} (around 1\% in mass fraction), about 6 orders of magnitude larger than the true value. The retrieval is most likely fitting the noise and/or spectral signatures of most of the other species at shorter wavelengths ($\lambda \lesssim 8\ \mu m$). Hence, when analyzing observations of potentially habitable terrestrial planets, we should be mindful not only of the false positive mechanisms that may be active in the atmosphere, but also of the false positives that the retrieval routines can infer. One could try to solve this issue by averaging over multiple retrieval runs, or by reducing the prior space with inferred knowledge from independent observations. 

The retrievals of the ancillary cloudy input spectra (shown in Appendix~\ref{app:cloudy}) do not show any noticeable improvement in either accuracy or precision for all scenarios with increased R and S/N. The values of $R_\mathrm{pl}$ and $T_0$ are still underestimated for all the scenarios. However, all considered R and S/N combinations still allow for an atmospheric abundance characterization for the cloudy input spectra. This analysis is subject to similar limitations as those already discussed for the subset of clear input spectra. The impact of clouds in retrievals will be discussed in more detail in Section~\ref{sec:clouds}.

\section{Discussion}\label{sec:discussion}

In Section~\ref{sec:CompIII} we compare the results we obtain for the cloud-free Modern Earth twin with the results from the similar study performed in Paper III. As previously mentioned, we retrieved spectra of cloudy exoplanets, while neglecting clouds in the forward model of the retrieval framework. We describe this effect in Section~\ref{sec:clouds}.
We discuss the impact of the quality of the data on the retrievals in Section~\ref{sec:quality} and the systematic effects for the retrieval runs in Section~\ref{sec:systematics}. Finally, we quantify the potential that \emph{LIFE} has in differentiating the various epochs (Section~\ref{sec:differentiating}).

For completeness, we mention that we also tested the impact of varying complexity of the theoretical spectral model on retrievals, by including and excluding scattering and/or CIA in the calculation. Through the analysis and comparison of ancillary retrieval grids, we could confirm that including or neglecting scattering and CIA in the calculation does not influence the quality of the results. We show the results in Appendix \ref{app:bayes}.

\subsection{Comparison with Paper III}\label{sec:CompIII}

\begin{figure}
   \centering
    \includegraphics[width=\linewidth]{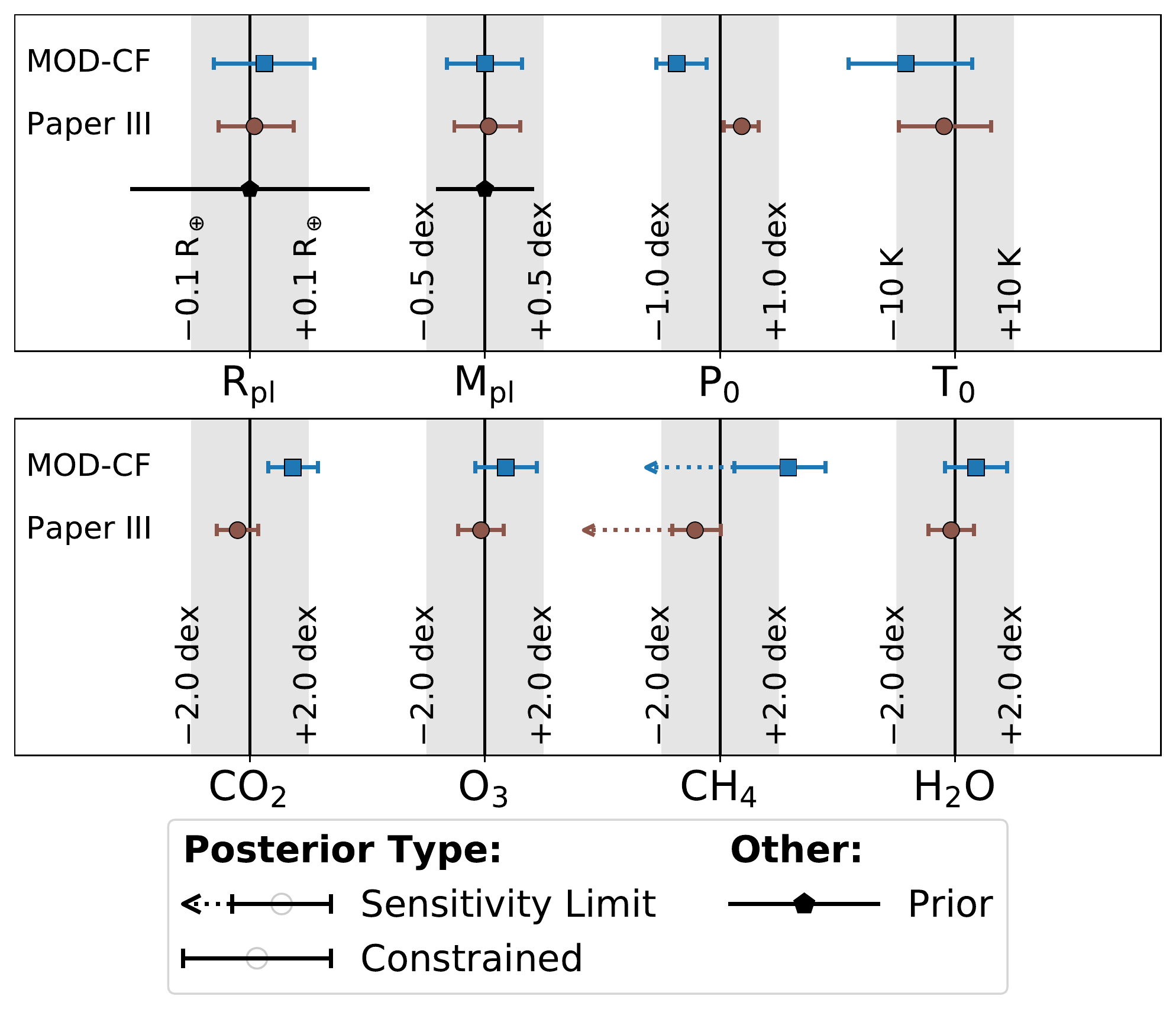}
   \caption{Comparison of retrieval results for constrained planetary parameters and atmospheric abundances in the clear Modern Earth (MOD-CF) case (blue, square marker) with results from Paper III (brown, circular marker) for input spectra with the same properties (wavelength coverage $4-18.5\,\mu$m, R~$=50$, S/N~$=10$). The vertical, black lines indicate the true values assumed for the parameters in each study. For parameters where we assumed a non-flat prior we indicate the prior range (black, pentagonal marker). The error bars on the Constrained posteriors denote the 68\% confidence intervals. }
    \label{fig:CompPaperIII}
\end{figure}

To allow for a proper comparison, we selected the model from Paper III that uses the same R, S/N, and wavelength range ($4-18.5\,\mu$m, R~$=50$, and S/N~$=10$). The major difference between these two retrieval studies is that in Paper III retrievals were performed using the same theoretical atmospheric model that was also used to generate the input spectra. In contrast, in this work, we used an atmospheric model in the retrieval which is different than the one that was used to generate the spectrum. Further, in our previous study we assumed abundance profiles that were vertically constant while the spectra calculated by \citet{Rugheimer2018} were based on a self-consistent, altitude-dependent atmospheric composition. In Figure~\ref{fig:CompPaperIII}, we compare the retrieval results for the constrained planetary parameters and abundances from Paper III to our findings for the clear Modern Earth case.

In the upper panel of Figure~\ref{fig:CompPaperIII}, we plot the retrieval results for the planetary parameters. The planetary radius $R_{\mathrm{pl}}$ is well constrained with respect to the assumed prior distribution and both posteriors are roughly centered on the corresponding truths. However, the spread of the clear Modern Earth $R_{\mathrm{pl}}$ posterior is larger than in Paper III, which indicates that the radius is slightly less well constrained. For $M_{\mathrm{pl}}$ our results are comparable to Paper III. 
Our results for the surface pressure $P_0$ and surface temperature $T_0$ agree less well with the results presented in Paper III. This is probably caused by small differences in the input P-T profiles, as well as potential systematic errors, which we will discuss in more detail in Section~\ref{sec:systematics}. 

In the lower panel of Figure~\ref{fig:CompPaperIII}, we show the results obtained for the abundances of the trace gases that were constrained (C- or SL-type posteriors) by our retrieval analysis. We observe that the clear Modern Earth retrieval tends to overestimate the true abundances, while the estimates from Paper III appear more accurate. The retrieved posterior types match for all of the atmospheric gases considered. Additionally, for \ce{CO2}, \ce{O3}, and \ce{H2O}, the spread of the posteriors for the clear Modern Earth runs are comparable to the results from Paper III. The larger spread in our \ce{CH4} abundance is the result of a slightly reduced sensitivity, which is most likely evoked by differences in the atmospheric scenarios used to generate the input spectrum and in the retrievals. These differences will reduce the accuracy overall of the retrieval results.

\subsection{Impact of clouds on retrieval results}\label{sec:clouds}

As pointed out in Section~\ref{sec:results}, using a cloud-free atmospheric model in our retrievals will likely have introduced biases into our results for the cloudy scenarios.

The presence of clouds in an atmosphere will reduce the MIR continuum emission of the observed exoplanet. The emission spectra of terrestrial exoplanets are typically dominated by the lowest, non-opaque atmospheric layers. In cloudy exoplanets, part of this thermal emission is hidden below the clouds. Thus, the atmospheric layers above the clouds contribute more to the overall spectrum. 
Because the atmospheric temperature at the top of the clouds is typically lower than the surface temperature of the planet, cloud coverage will generally lead to a cooler retrieval temperature with reduced continuum flux. This reduction in continuum flux can be clearly seen when comparing the clear to the cloudy spectra in Figure~\ref{fig:RetSpectra}.

Since the theoretical spectral model we use for our retrievals assumes a cloud-free atmosphere, the continuum emission must be reduced in other ways to achieve a satisfactory fit to the input spectrum. This reduction can be obtained by reducing the radius (and thus the emitting surface) and/or the surface temperature of the exoplanet. Due to these compensation effects, most of our retrievals of cloudy input spectra yielded smaller radii and/or cooler surface temperatures than the cloud-free inputs (see Figure~\ref{fig:parameters_epochs}).  This is also valid at higher spectral resolutions and signal-to-noise ratios, showing that the compensation effects are independent from the quality of the data (see Figure~\ref{fig:QualityParamsCloudy}).

In addition to the surface conditions, the thermal structure of the layers and the chemical composition of the atmosphere also play a major role in shaping the emission spectrum, especially in the absorption/emission features. This could yield biased retrieval results for other parameters, such as the $a_i$ coefficients of the polynomial P-T profile. In our results, the cloudy Prebiotic Earth (PRE-C) model shows the clearest signature of this compensation effect (see Figure~\ref{fig:RetProfiles}). This degeneracy between cloud coverage and thermal structure in retrievals was also found in other studies \citep[see e.g.][and references therein]{2020A&A...640A.131M}.

Such biased results could be misleading, especially when trying to analyze the habitability of an observed exoplanet. If, by neglecting clouds in our theoretical spectral model, we underestimate the surface temperature, we could therefore misclassify habitable exoplanets. A clear example is the cloudy Modern Earth (MOD-C) scenario: using our cloud-free forward model to retrieve this cloudy spectrum causes the retrieved ground temperature to be colder than 275 K. Such low temperatures suggest a potentially uninhabitable planet, which we know is not the correct interpretation for the cloudy Modern Earth spectrum.

On the other hand, when looking at the retrieved chemical abundances (see Figure~\ref{fig:abundances_epochs}), we observe only minor variations in the shape of the posteriors for all the major absorbing gases in the atmosphere. This indicates that, despite having a major impact on the retrieved physical parameters ($R_\mathrm{pl}$ and $T_0$), retrieving cloudy spectra with a cloud-free model does not significantly impact the chemical characterization of the atmosphere.

Therefore, including a cloud model in the theoretical spectral model that we use for retrievals could improve the quality of the results.  However, this depends on the goal of the analysis. If we aim to characterize the chemical composition of the atmosphere, it may be sufficient to use cloud-free retrievals. This would be a smart strategy considering that including a somewhat realistic cloud treatment in the theoretical spectral model significantly increases the number of retrieved parameters and subsequently the running time. 
Performing retrievals on input spectra that include visible/near-infrared data in addition to the MIR observations will likely provide additional information about the cloud composition and structure. In this sense, coupling with data acquired by HabEx/LUVOIR and \textit{LIFE} may significantly improve retrieval results. We will compare the retrieval performance for different cloud models and discuss the capabilities of joint reflected light/thermal emission retrievals in future publications.

\subsection{Increasing the quality of the input spectra}\label{sec:quality}

The results of the retrievals performed assuming other combinations of R and S/N, described in Section~\ref{sec:r-snr}, show that increasing the S/N to 20 will allow us to detect both \ce{O3} and \ce{CH4} in more cases. Increasing to R~$=100$ would also improve our results, especially when combined with an increase in S/N. This is an interesting finding for multiple reasons. 

From the scientific point of view, simultaneously detecting \ce{O3} (which can provide an indirect estimate of \ce{O2}) and \ce{CH4} would be a strong indicator of chemical disequilibrium in the atmosphere possibly hinting at the existence of biological activity. Such a detection would make the respective exoplanet a high-priority target for the search of life beyond the Solar System. This concept will be further explored in Section~\ref{sec:differentiating}. 

From the technical point of view, it would mean that one needs to consider longer integration times, while maintaining a stable architecture of the interferometer array. For the assumptions in our baseline case (see, Table~\ref{tab:lifesim}),   
doubling the resolution would roughly correspond to a doubling of the integration time (from $\sim50$ to $\sim100$ days),  while doubling the S/N would translate in integration times roughly four times longer (from $\sim50$ to $\sim200$ days). This poses challenges in terms of mission technical feasibility as well as mission scheduling. Increasing the instrument throughput, for which we assumed a conservative value (cf. Paper I), or the aperture size would bring the required integration times down. Also, the nearest rocky exoplanets orbiting within the habitable zone (HZ) of their Solar-type host stars may not be 10 pc away. \cite{Bryson2021} estimated that with 95\% confidence the nearest HZ planet around G and K dwarfs is $\sim$6 pc away and they predict $\sim$4 HZ rocky planets around G and K dwarfs within 10 pc of the Sun. Taking all this together, we would therefore recommend the stick to the baseline requirements for \emph{LIFE} of R~$=50$ and S/N~$=10$, as proposed in Paper III, since they allow for a reliable and quantitative characterization of the most important physical and chemical properties of the considered atmospheres. The most promising targets could then be observed further to increase the S/N, thus allowing a more precise characterization of the atmosphere.

\subsection{Systematics and current challenges}\label{sec:systematics}

Thus far, we can confidently conclude that our Bayesian framework can retrieve consistent and robust results. This is not only valid for simulated observations generated with \texttt{petitRADTRANS} (see Paper III), but also for input spectra produced by other radiative transfer models \citep[here by][]{Rugheimer2018}. These results are highly promising in the context of analyzing real observational data in the future. However, as we mentioned in the previous sections, our work has identified some aspects which may lead to biased results. 
Some issues are linked to the intrinsic limitations of the Bayesian retrieval routine we described in Section~\ref{sec:limitations}. Ideally, these can be mitigated to improve the results, for example by choosing a different P-T profile parametrization, or by adding a cloud model to the retrieval. Further, we purposely chose to perform our retrievals assuming uniform priors for most parameters where all values were possible if within a specified, wide range (see Section~\ref{sec:methods} for details). However, for future observations, the prior space might already be constrained (e.g. if one or more parameters are already measured by independent observations) and this would likely improve our retrieval results. 

Despite these possibilities, we will eventually be limited by two factors. First, the number of parameters that the Bayesian framework can handle within reasonable computing time. This limit on the number of parameters will remain unless novel parameter estimation algorithms emerge. An example would be the use of machine-learning retrieval routines \citep[e.g.][]{2016ApJ...820..107W,2018NatAs...2..719M,2019AJ....158...33C}. Second, for a given resolution the information content of the spectrum is limited. Therefore, considering additional parameters in the retrieval framework could bias the results, for example causing a false positive inference of an atmospheric species.

\begin{figure*}
    \centering
    \includegraphics[width=0.95\linewidth]{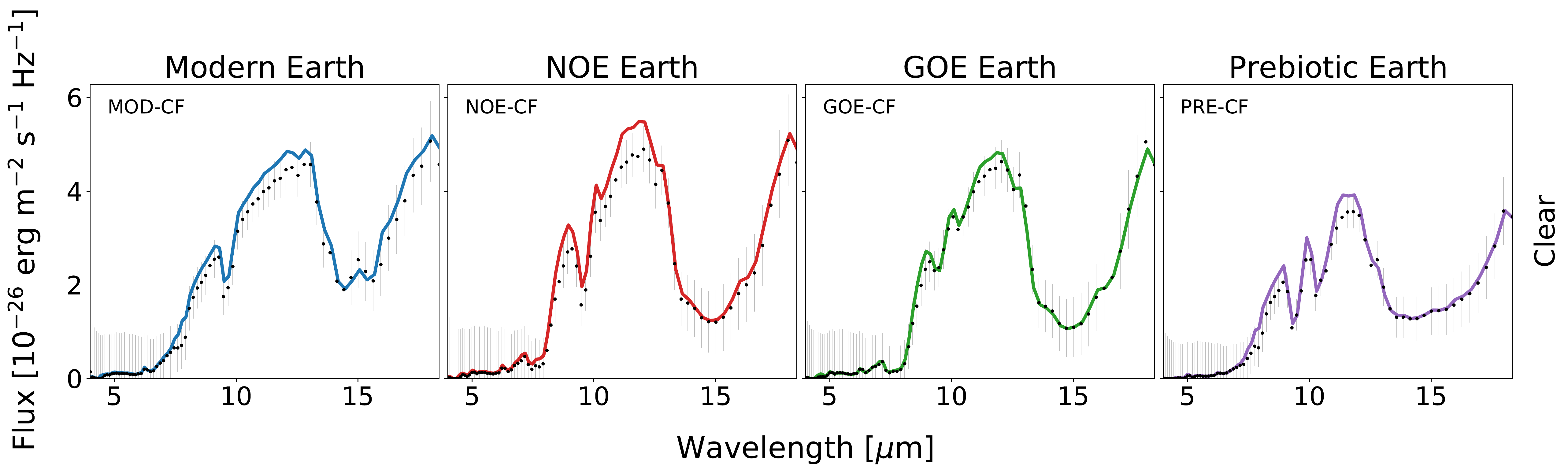}
    \caption{Comparison between the R~$=50$ MIR spectra of the four clear-sky epochs (MOD-CF, NOE-CF, GOE-CF, and PRE-CF;  solid lines following the color scheme of Table~\ref{table:id}) produced with \texttt{petitRADTRANS} with the results from \citet{Rugheimer2018} (black dots) assuming the same input parameters (i.e. P-T profile, abundances, planetary dimensions).  The error bars indicate the LIFE\textsc{sim} uncertainty assumed for the main grid of retrievals (S/N~$=10$ at 11.2 $\mu$m).}
    \label{fig:spectra_systematics}
\end{figure*}

However, the most relevant issues are independent of the parameter estimation routine. They are rooted in the intrinsic differences between individual radiative transfer models used to produce the MIR input spectra and the theoretical spectra in the retrievals. Such discrepancies may be caused by a slightly different treatment of physical or chemical processes, or differences in the assumed opacity tables. 
To investigate these issues, we computed the MIR spectra for the four clear-sky scenarios (MOD-CF, NOE-CF, GOE-CF, and PRE-CF) using \texttt{petitRADTRANS}. We assumed exactly the same input parameters (i.e. P-T profile, abundances, planetary dimensions) that \citet{Rugheimer2018} used to produce their spectra. We show the results for R~$=50$ in Figure~\ref{fig:spectra_systematics}. The \texttt{petitRADTRANS} spectra are plotted as solid lines using the color scheme from Table~\ref{table:id}. The input spectra from \citet{Rugheimer2018} are shown as black dots. The error bars indicate the LIFE\textsc{sim} uncertainty used in the main grid of retrievals (S/N~$=10$ at 11.2 $\mu$m). 

We observe that the \texttt{petitRADTRANS} spectra deviate (mostly within the LIFE\textsc{sim} uncertainty) from the spectra calculated by \citet{Rugheimer2018}, despite both models assuming the exact same input. While the absorption features are generally in agreement with each other, the spectra produced by \texttt{petitRADTRANS} show a higher continuum flux, especially around $8-12$ $\mu$m. This discrepancy is likely linked to differences in the opacity tables used by the two radiative transfer models. As stated in Section~\ref{sec:limitations}, these differ with respect to:

\begin{enumerate}
 \item \emph{Wing Cutoff}. To prevent the wings of the pressure-broadened lines from extending to infinity (non-physical), it is necessary to introduce a wing cutoff. However, different radiative transfer models assume different cutoff thresholds \citep[see the comparisons performed by e.g.][]{2019MNRAS.487.2082L,2017ApJ...850..150B,Barstow2020}. \citet{Rugheimer2018} used a wing cutoff at 25 $\mathrm{cm^{-1}}$ from the line center. In contrast, the line cutoff used for the \texttt{petitRADTRANS} opacity tables assumes an exponential line wing decrease \citep[for details see][]{2019A...627A..67M}. This may explain the higher continuum emission observed for all \texttt{petitRADTRANS} spectra.
    \item \emph{Line List Databases}. The default opacity tables used by \texttt{petitRADTRANS} stem from different sources. They are calculated from the HITEMP, HITRAN 2012, or ExoMol line lists (see Table~\ref{table:opacities}). In contrast, the spectra from \citet{Rugheimer2018} were computed using only the HITRAN 2016 line lists, which in some cases are more recent than the ones adopted in our study. At the pressures and temperatures of interest in the study, we wouldn't expect large variations in the line lists, provided all the databases are synchronous. The use of different versions of the same database (e.g. HITRAN 2012 versus HITRAN 2016) might cause variations in the opacities, since databases more recently updated generally include more transition lines \citep[see e.g.][]{2017JQSRT.203....3G}. Furthermore, the default \texttt{petitRADTRANS} opacities only account for transitions of the main isotope, while the opacity tables used in \citet{Rugheimer2018} can account for additional isotopes.
     \item \emph{Pressure Broadening Coefficients}. To compute the line profiles, it is necessary to account for collision-driven line broadening. This depends on the pressure and composition of each atmospheric layer. For most molecules both models assume air broadening, which is based on a Modern Earth-like atmospheric composition. However, for \ce{CH4}, the \texttt{petitRADTRANS} opacity table assumes a theoretical broadening model based on Equation 15 in  \citet{2007ApJS..168..140S}, which was experimentally validated. Another exception is \ce{N2O}, for which H-He broadening was assumed \citep[see ][]{2021A...646A..21C}. However, at the pressures and temperatures of interest, we do not expect large differences due to pressure broadening \citep{2007ApJS..168..140S,2019A...627A..67M,2019ApJ...872...27G,2021A...646A..21C}. We mention it here for completeness.
    
\end{enumerate}

These differences likely also account for a substantial part of the offsets we find in the retrieved parameter values.  Future inter-comparison studies could help us define a ``best practice'' upon which to agree, as a community, to compute opacity tables for retrievals in order to minimize these systematic effects. Furthermore, an ongoing experimental work will be necessary to improve the completeness of the transition line databases and reduce discrepancies.

\subsection{Differentiating the epochs}\label{sec:differentiating}

A quantitative approach to differentiate between the various scenarios is through the results of the retrieval analyses. We performed a first qualitative step in this direction in Sections \ref{sec:ptprofiles}, \ref{sec:planetaryparameters}, and \ref{sec:abundances}, where we visually compared the retrieved P-T profiles and the posteriors for the planetary parameters and abundances. Through visual comparison, we found that differentiating the epochs via the retrieved P-T structure and planetary parameters is  challenging. By studying the retrieved abundance posteriors, we found that the best candidates to perform such differentiation are \ce{O3} and \ce{CH4}. This finding is especially interesting since the \ce{O2}-\ce{CH4} pair is generally considered the strongest biosignature \citep[see][]{Lovelock:CH4-O2,1965Natur.207....9L} and \ce{O2} can be constrained from \ce{O3} through atmospheric chemistry models. Thus, the detection of one or both of these molecules will likely trigger follow up observations and could allow us to separate between potentially alive and lifeless planets. However, a more in-depth characterization of the atmospheres is limited by the large variance on the posteriors of all these species, which typically exceeds one order of magnitude. 

A more quantitative separation between the retrieved posterior distributions for the various epochs can be achieved by considering the difference between the cumulative posterior distribution functions of two epochs for a model parameter. This approach is similar to the Kolmogorov-Smirnov test \citep{kolmogorov1933sulla,smirnov1939estimation}, which is generally used to assess whether two samples are drawn from the same underlying distribution. Given a model parameter $M$ with prior range $X=[X_\mathrm{min},X_\mathrm{max}]$, we calculate the cumulative distribution $G^M(x)$ for $x\in X$ of the retrieved posterior $P(x)$ as follows:
\begin{equation}
    G^M(x)=\int_{X_\mathrm{min}}^x P\left(x'\right) dx'\cdot\left(\int_{X_\mathrm{min}}^{X_\mathrm{max}} P\left(x'\right) dx'\right)^{-1}.
\end{equation}
We then compare the cumulative distribution functions $G_a^M(x)$ and $G_b^M(x)$ of two different epochs $a$ and $b$, by considering the maximum difference $\Delta := \Delta^M_{a-b}\in[0,1]$ between them:
\begin{equation}
    \Delta = \max{|G_a^M(x)-G_b^M(x)|}.
\end{equation}
Thus, small values of $\Delta$ indicate that the compared posterior distributions only show small differences relative to each other. In this case it is hard to differentiate between the retrieved posteriors. On the other hand, larger values of $\Delta$ indicate that the differences between the two posteriors are likely to correspond to different underlying true values of the considered parameter. 

\begin{figure*}
 \centering
    \subfloat{{\includegraphics[width=0.47\linewidth]{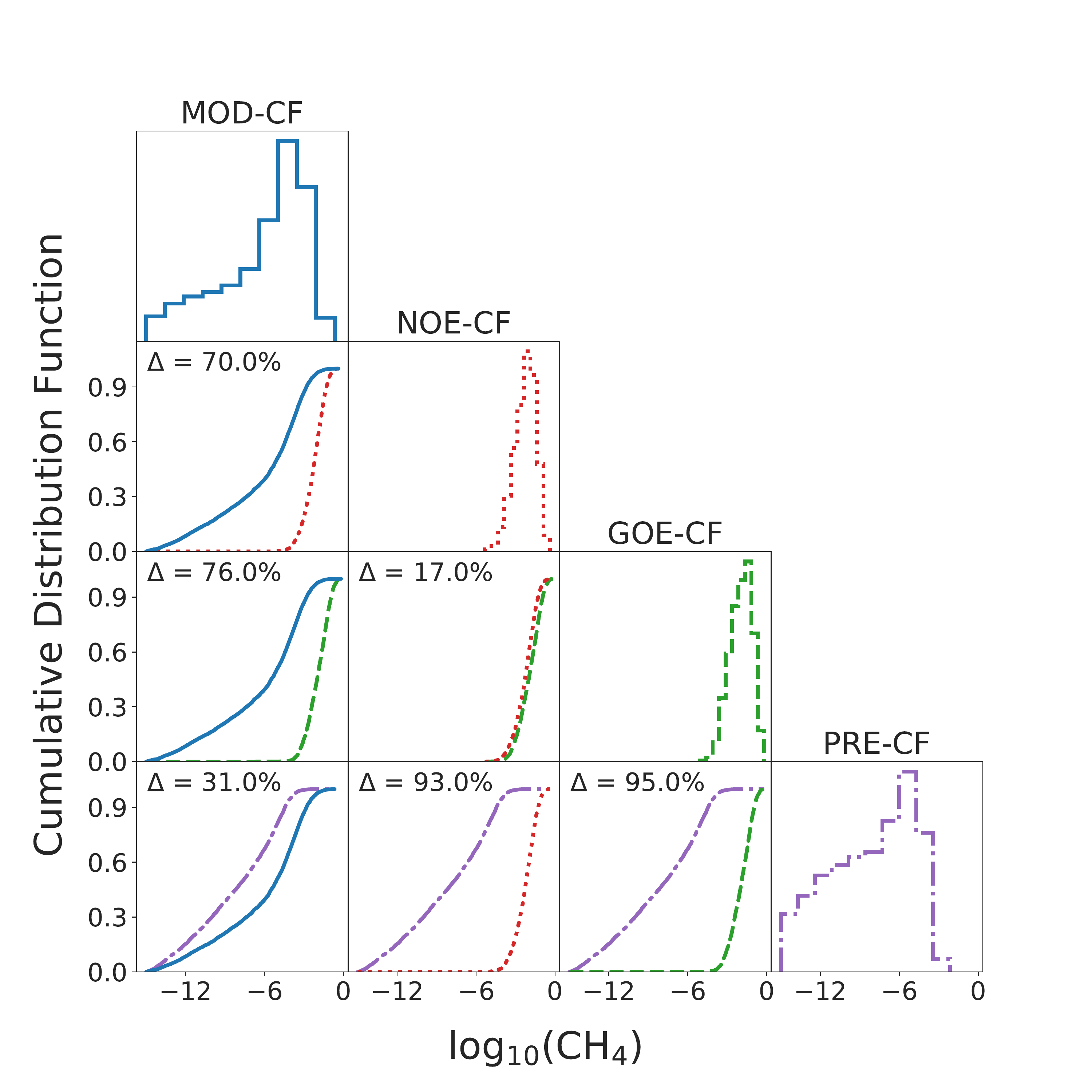} }}%
    \qquad
    \subfloat{{\includegraphics[width=0.47\linewidth]{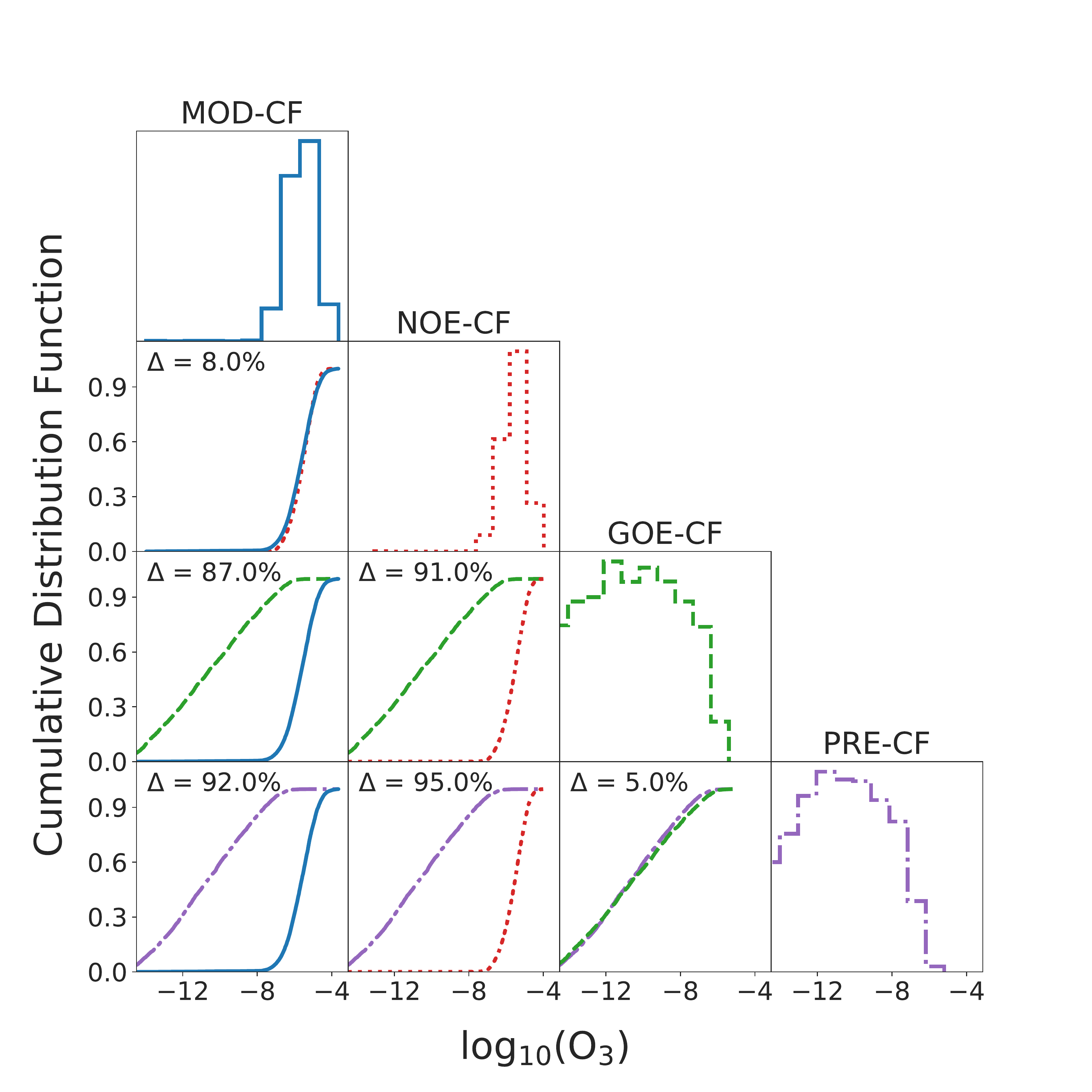} }}%
    \caption{\emph{Left:} Comparison of the cumulative distribution functions of the \ce{CH4} posteriors for all the combinations of the clear sky scenarios (MOD-CF, NOE-CF, GOE-CF, PRE-CF). The retrieved \ce{CH4} posteriors for each scenario are shown on the diagonal. Following the color scheme in Table~\ref{table:id}, we show the posteriors and cumulative distribution functions as: solid blue lines (MOD-CF); dotted red lines (NOE-CF); dashed green lines (GOE-CF); dash-dotted purple lines (PRE-B).  \emph{Right:} Same plot, but for \ce{O3}.}%
    \label{fig:example}%

\end{figure*}
We can calculate $\Delta$ for all the combinations of the various scenarios and for all parameters. We get particularly interesting results for \ce{CH4} and \ce{O3}.
Figure~\ref{fig:example} shows the cumulative distribution functions for all the combinations of the clear sky scenarios (MOD-CF, NOE-CF, GOE-CF, PRE-CF) calculated from the posteriors of \ce{CH4} and \ce{O3}, for R~$=50$ and S/N~$=10$. Annotated in each subplot of the corner plot, we noted the values of $\Delta$ (percentage) corresponding to each combination. On the diagonals, the retrieved posteriors for every scenario are shown for reference. We keep the color scheme defined by Table~\ref{table:id}.
Regarding \ce{CH4}, we can fairly confidently distinguish between the clear prebiotic Earth (PRE-CF) and the Earth after the GOE (GOE-CF), for which $\Delta=95\%$, as well as between PRE-CF and the Earth after the NOE (NOE-CF), for which $\Delta=90\%$. The distinction between the prebiotic Earth and the Modern Earth (MOD-CF), as well as between the NOE and the GOE Earth is more difficult ($\Delta\leq31\%$). For \ce{O3}, we observe a clear division into two subgroups: on the one side the Modern and NOE Earth, where we have a clear detection of \ce{O3}, and on the other hand the GOE and prebiotic Earth, where we only retrieve an upper limit on the abundance. The high value of $\Delta\sim 90\%$ between all combinations of MOD-CF/NOE-CF versus GOE-CF/PRE-CF clearly allows such a distinction. 
This is in agreement with what we concluded from in Figure~\ref{fig:abundances_epochs} in Section~\ref{sec:abundances}. However, in contrast to the qualitative discussion based on the posteriors appearance, $\Delta$ provides a promising metric to quantify the magnitude of these differences.

\begin{figure*}
    \centering
    \includegraphics[width=\linewidth]{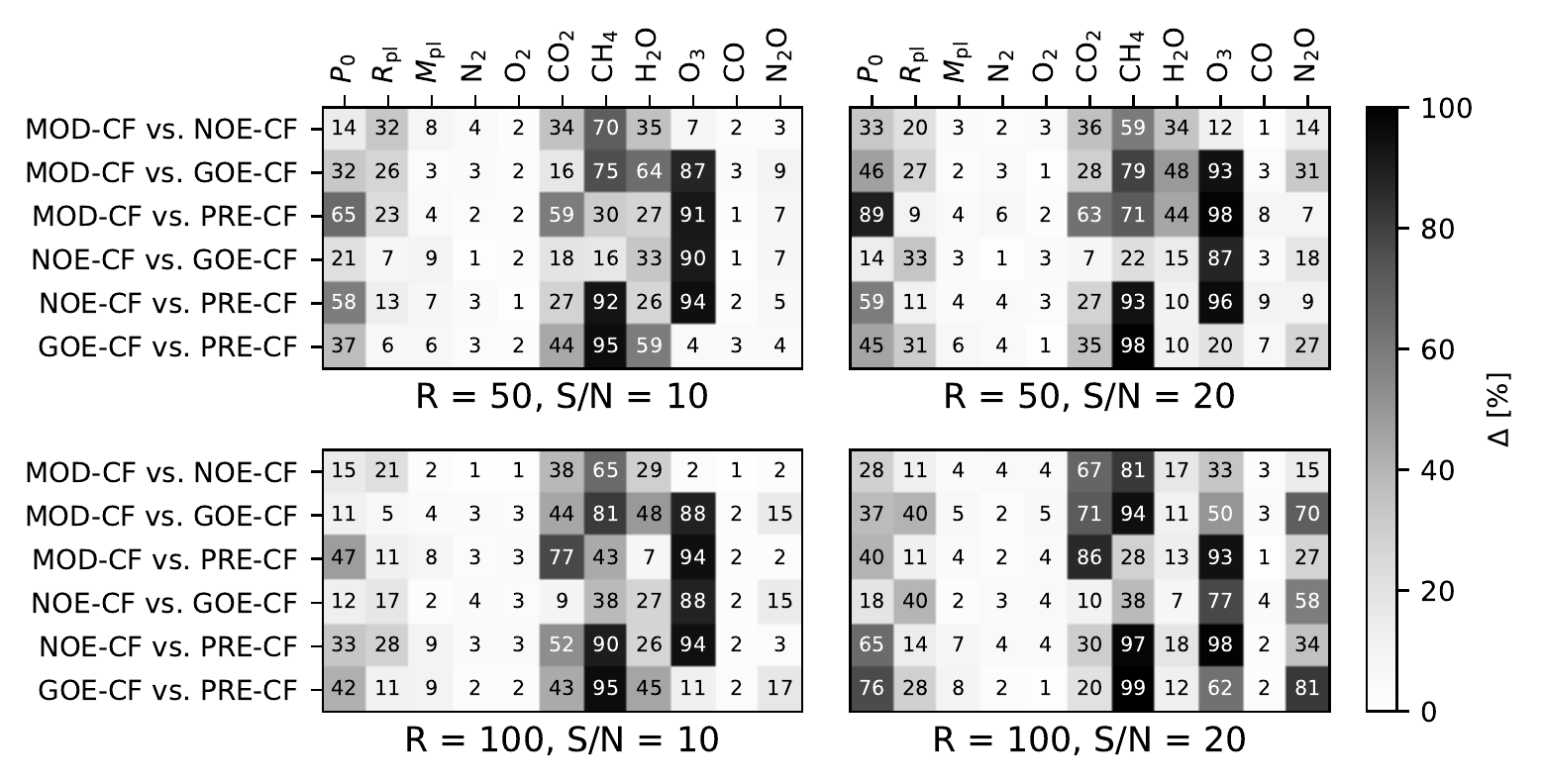}
    \caption{Maximum difference $\Delta$ between the cumulative posteriors for the different model parameters, for each  combination of input spectra (cloud-free subset), and different R-S/N pairs. The background of each cell in the tables is related to the value of $\Delta$ (darker hues for larger $\Delta$). }
    \label{fig:DeltaAll}
\end{figure*}

In Figure~\ref{fig:DeltaAll} we summarize the calculated $\Delta$ values for all combinations of cloud-free input spectra and for the different R-S/N pairs considered in Section~\ref{sec:quality}, for a total of four tables. Within the tables, each cell shows the $\Delta$ value (percentage) for a given parameter (columns) and for a comparison of two specific scenarios (rows). The cells are also colored according to the value of $\Delta$, with darker hues for larger values of $\Delta$.
As mentioned above, the biggest differences between the posteriors at R~$=50$, S/N~$=10$ can be observed for the molecules \ce{CH4} and \ce{O3}. Furthermore, we observe some differences for \ce{CO2} and \ce{H2O}, as well as for the $P_0$ posteriors. However, as discussed previously, these differences are rooted in degeneracies between the pressure and the abundances (see Section~\ref{sec:CompIII}) and not  caused by large physical differences in the underlying atmospheres. These findings are generally still valid as we move to higher R and S/N. Similar conclusions can also be drawn for the cloudy inputs (see Appendix~\ref{app:cloudy}).

Since we are able to confidently detect \ce{O3} for a clear Earth after the NOE and for the Modern Earth and we can distiguish these two epochs from earlier scenarios (Prebiotic and GOE Earth), we can infer that \emph{LIFE} would be able to detect traces of life as we know it in an Earth-like atmosphere when the abundance of \ce{O2} has passed the 10\% PAL threshold. This is consistent with other studies that focused on different wavelength ranges, such as the work by \citet{2019AJ....157..213K} based on LUVOIR.
The biosignature pair \ce{CH4}-\ce{O3} might be even easier to detect when the abundance of \ce{O2} is around 10\% PAL (NOE Earth), rather than Modern Earth. The NOE Earth scenario is particularly favored since the atmosphere is filled with enough \ce{O3} to be detectable, but a low enough abundance of \ce{O2} to deplete the \ce{CH4} in the atmosphere. These results are also consistent with the results shown in \citet{2019AJ....157..213K}.
In other words, if \emph{LIFE} were to observe the Earth at various stages of its evolution orbiting the Sun at 10 pc distance, it would be able to detect strong indicators of life starting from around 0.8 Ga (NOE Earth). The detection of \ce{CH4} with an upper limit on \ce{O3} would also allow a tentative detection of potential biological activity up to 2.0 Ga (GOE Earth).

We must keep in mind that the epochs that we chose for our study are momentary "snapshots" in the continuous evolution of Earth, even though these four scenarios represent the major changes in our atmosphere. Still, other evolutionary paths are possible in the context of exoplanets, especially when considering other stellar classes. Realistically, all promising candidates would be followed-up with additional observations within the \textit{LIFE} mission. 
It is beyond the scope of this work to conclusively infer the presence of a biosphere from the measured spectra of potentially habitable candidates, As discussed in other works such as \citet{2018AsBio..18..630M} or \citet{KriTo2022_context}, we would require a thorough discussion of the context information available for the observed planetary system before claiming a ``life detection''. However, the presented retrieval results are certainly an important piece of information for the  development of frameworks for systematically  assessing biosignature detections \citep[e.g.][]{Cat2018Framework,Walk2018framework}.

\section{Conclusions}\label{sec:conclusions}

 The Bayesian retrieval framework introduced in Paper III and extended here has delivered insightful answers to the questions introduced in Section~\ref{sec:intro}. These can be summarized as follows:
\begin{enumerate}
    \item 
    \emph{LIFE} can characterize pre- and biotic worlds. We can constrain the surface temperatures with an uncertainty of around 20 K. We can confirm, exclude or give upper limits on the presence of several astrobiologically relevant molecules that show signatures in the MIR bands. In particular, \emph{LIFE} could be able to detect \ce{O3} in the atmosphere if the \ce{O2} mass fraction is of the order of $\sim$10\% PAL. \ce{CH4} could be constrained in terrestrial atmospheres if its abundance is $\sim0.1\%$ in mass fraction. For lower abundances (around $10^{-6}$ in mass fraction), \emph{LIFE} would detect an upper limit on \ce{CH4} (SL-type posterior). Simultaneously constraining \ce{O3} and \ce{CH4} would be possible in atmospheres with an abundance of \ce{O2} around 10\% PAL. This is in agreement with other studies based on a different wavelength range \citep{2019AJ....157..213K}.  Such result is relevant in terms of the detection of biosignatures in the atmospheres of habitable exoplanets. 
    \item 
    Neglecting clouds in the retrievals could cause biases in the determination of the thermal structure of the atmosphere of cloudy exoplanets. However, cloud-free retrievals of cloudy spectra would still yield accurate results for what concerns the atmospheric composition.

    \item 
     We confirmed that the minimum requirements in spectral resolution and S/N for an MIR mission like \emph{LIFE} found in Paper III are also sufficient for the scenarios considered here. Improving the S/N would allow for a clearer detection of \ce{O3} and \ce{CH4} even when these species are less abundant (until $\sim10^{-7}$ in mass fraction). Therefore, a better characterization could be obtained by observing promising targets longer during the characterization phase of the mission (or by increasing the instrument throughput and/or aperture size).
    \item 
    We were able to demonstrate that inter-model comparison and retrieval is possible with the caveats and limitations detailed in Section~\ref{sec:systematics}. The most important discrepancies in the retrievals are caused by the use of different opacity tables, in particular for what concerns the line wing cutoff treatment. Degeneracies and correlations in the posteriors appear as a result of the various relations among various parameters. 
\end{enumerate}

\section{Next Steps}\label{sec:future}

Several new interesting questions and opportunities for more detailed studies arise from this work. First of all, we plan to work on a study that will take advantage of the model selection potential that Bayesian retrievals have to offer, for example by comparing retrievals including and excluding non-retrieved parameters (e.g. the CO abundance). We are also performing retrievals assuming various cloud models (Konrad et al., in prep.). Retrievals of hazy planets \citep[see e.g.][]{2016AsBio..16..873A}, as well as ocean worlds, might also help us further quantify the science potential of \emph{LIFE} for a variety of different planet types.

Another interesting study would be to increase S/N and R to even higher values. This will not only evaluate the extreme limits of a concept like \emph{LIFE}, but also help us better understand if retrievals are limited by R rather than S/N (e.g. due to unresolved narrow features at low R). It would also be useful to compare different R-S/N combinations, this time fixing the observing time. This would help us quantify the best R-S/N combination needed to optimize the characterization of a terrestrial atmosphere. 
Further work is needed to optimize the yield in the characterization phase of the \emph{LIFE} mission concept. The estimates of the observation time needed to establish knowledge about the habitability and the presence of biologically relevant molecules in the atmosphere that we derived here are a crucial piece of information for these follow-up studies.

In this work, we only used simulated data obtained with the \emph{LIFE} mission. However, in the future there will likely be more information available to each system and planet. Therefore, it will be important to put this study in context with other observations. For instance, joint retrievals of reflected light data obtained with LUVOIR/HabEx at optical/near-infrared wavelengths and thermal emission spectra as obtained by \emph{LIFE} would provide useful insight on the synergies between the various missions.

One of the most important open questions regarding the ultimate goal of detecting extrasolar life will require to put our results in context with life detection frameworks \citep[e.g.][]{2021Natur.598..575G,2018AsBio..18..709C,2018AsBio..18..779W}. Our ongoing retrieval efforts could be useful for the fine-tuning of such frameworks. These, in turn, would provide insight on the meaning and the likelihood of a potential biosignature detection, which would allow us to infer and justify the presence of life forms on another planet.

\begin{acknowledgements}
      This  work  has  been carried  out within  the  framework  of the National Center of Competence in Research PlanetS supported by the Swiss National  Science  Foundation. S.P.Q. and E.A. acknowledge the financial support from the SNSF. P.M. acknowledges support from the European Research Council under the European Union’s Horizon 2020 research and innovation program under grant agreement No. 832428. J.L.G. thanks ISSI Team 464 for useful discussions.
      \\
      \\
        \emph{Author contributions.} E.A. carried out the analyses, created the figures, and wrote the bulk part of the manuscript. B.S.K. and D.A. wrote part of the manuscript. S.P.Q. initiated the project, guided the project and wrote part of the manuscript. All authors discussed the results and commented on the manuscript.
        \\
        \\

      \emph{Software.} This research made use of: Astropy\footnote{http://www.astropy.org}, a community-developed core Python package for Astronomy \citep{astropy:2013, astropy:2018}; Matplotlib\footnote{https://matplotlib.org/3.1.1/index.html} \citep{Hunter:2007}; pandas \citep{reback2020pandas}; seaborn \footnote{https://seaborn.pydata.org}.
      
\end{acknowledgements}

%
%
\typeout{}
\bibliographystyle{aa}
\bibliography{biblio} 
%

%

%


\begin{appendix} 
\section{Scattering of terrestrial exoplanets}\label{app:scattering}

As discussed in \citet{2020A&A...640A.131M}, \texttt{petitRADTRANS} was updated to treat scattering. This was done using the Feautrier method \citep{1964CR....258.3189F}.
This is a third-order method that allows the treatment of the radiative transfer equation in the diffusive regime. 

The Feautrier method solves the angle- and frequency-dependent radiative transfer equation for both the planetary and the stellar radiation field. These can be treated separately, since the radiative transfer equation (Eq. \ref{eq:radtrans}) depends only linearly from the intensity $I$. \begin{equation}
\mu \frac{dI}{d\tau}=-I+S.\label{eq:radtrans}
\end{equation}
Here, $\mu=\cos{\theta}$ where $\theta$ is the angle between a light ray and the surface normal, $\tau$ is the optical depth, $I$ is the intensity, and $S$ is the source function.  

Conceptually, for any direction $\mu$ of a ray, there also exists a ray in direction $-\mu$, where $\mu \in [-1,1]$. It is possible to instead let $\mu$ run from 0 to 1 only, and define rays $I_+$ and $I_-$ parallel and anti-parallel to this direction. For one of these, the projection onto the atmospheric normal vector (defined by the scalar product) will be positive (going upward), while for the other one it will be negative (that is, going downward). Eq. \ref{eq:radtrans} can be therefore rewritten as:

\begin{equation}
    \frac{dI_+}{d\tau}=S-I_+\label{eq:Iplus}
\end{equation}

\begin{equation}
    \frac{dI_-}{d\tau}=-S+I_-.\label{eq:Iminus}
\end{equation}

To solve these, it is convenient to define other two variables:

\begin{equation}
    I_J=\frac{1}{2}(I_++I_-)
\end{equation}

\begin{equation}
    I_H=\frac{1}{2}(I_+-I_-)
\end{equation}

So that Eqs. \ref{eq:Iplus} and \ref{eq:Iminus} become:

\begin{equation}
    \frac{dI_J}{d\tau}=-I_H
    \label{eq:IJIH}
\end{equation}

\begin{equation}
    \frac{dI_H}{d\tau}=S-I_J.\label{eq:IHIJ}
\end{equation}

Replacing $I_H$ as defined by Eq. \ref{eq:IJIH} into \ref{eq:IHIJ}, we obtain Feautrier's equation:
\begin{equation}
    \frac{d^2I_J}{d\tau^2}=I_J-S.\label{eq:feautrier}
\end{equation}

In this paper, we  only take into account thermal scattering, i.e. scattering of the planetary radiation. We therefore neglect the scattering of the direct stellar contribution. However, since the radiative transfer equation depends only linearly on $I_\nu$, the contribution of the stellar radiation can be treated as an additional term in the calculation \citep[see][]{Molliere2017}. This term is also included in the latest version of \texttt{petitRADTRANS} and we refer to the online documentation for a more detailed description\footnote{ \url{https://petitradtrans.readthedocs.io/en/latest/content/notebooks/emis_scat.html\#Scattering-of-stellar-light}}. 

Purely considering the planetary radiation, we define the boundary conditions at the top of the atmosphere:

\begin{equation}
    I_+(P=0,\mu)=0\quad \forall \mu
\end{equation}
meaning that there is no planetary radiation coming downwards from the top of the atmosphere, and at the surface:

\begin{equation}
    I_-(P=P_{surf},\mu)=e_{surf}\ B(T_{surf})+a_{surf}J^{scat}(P_{surf}).
\end{equation}


The bottom boundary constrain on $I_-$ is composed by the thermal emission of the surface itself (blackbody radiation scaled by the surface emissivity $e_{surf}$) and by a portion of the incoming planetary radiation that is reflected by the surface. The wavelength dependence of the effectiveness of the reflection depends on the "surface albedo" or "reflectance" $a_{surf}$. 
The average scattered intensity $J^{scat}$ is the integral of $I_+$ over the all the possible angles ($\theta\in [-\frac{\pi}{2},\frac{\pi}{2}]$, which corresponds to the light that comes from the top layers):
\begin{equation}
    J^{scat} (P_{surf})= \int_0^1 I_+(P_{surf})d\mu
\end{equation}


The boundary conditions translate, in terms of $I_H$ and $I_J$, in:

\begin{equation}
    I_J(0) = -\frac{I_-(0)}{2}
\end{equation}
and


\begin{equation}
    I_J(P_{surf}) = \frac{1}{2}\left[I_+(P_{surf})+e_{surf}\ B(T_{surf})+a_{surf}J^{scat}(P_{surf})\right]
\label{eq:boundary}\end{equation}

It is possible to thus solve Eq. \ref{eq:feautrier} for $i\neq 1$, $i\neq N$ by discretization:

\begin{equation}
    -\frac{\left(\frac{I_{J,i+1}-I_{J,i}}{\tau_{i+1}-\tau_{i}}\right)-\left(\frac{I_{J,i}-I_{J,i-1}}{\tau_{i}-\tau_{i-1}}\right)}{\left(\frac{\tau_{i+1}+\tau{i}}{2}-\frac{\tau_{i}+\tau{i-1}}{2}\right)}+I_{J,i}=S_i
\end{equation}

Which can be expressed in matrix form by extracting the coefficients $a_i$, $b_i$, and $c_i$. 

\begin{equation}\renewcommand\arraystretch{1.6}
    \begin{pmatrix}
b_1 & c_1 & 0 & \cdots & 0\\
a_2& b_2 & c_2 &  \ddots & \cdots\\
0 & \ddots & \ddots & \ddots & 0\\
\vdots &\ddots &a_{N-1}& b_{N-1} & c_{N-1} \\
0 &\cdots &0 & a_N & b_N\\
\end{pmatrix}\cdot \begin{pmatrix}
I_{J,1} \\
I_{J,2}\\
\vdots\\
I_{J,N-1}\\
I_{J,N}\\
\end{pmatrix} = \begin{pmatrix}
S_{1} \\
S_{2}\\
\vdots\\
S_{N-1}\\
S_{N}\\
\end{pmatrix}
\end{equation}

To take into account the boundary conditions, at $i=1$ the value of $a_1$ is 0, while at $i=N$ both $c_N$ and $a_N$ will be 0, since there is no dependence from the $(N-1)$th layer in the boundary condition \ref{eq:boundary}; $b_N$, as a consequence, will be equal to 1.

The tridiagonal matrix can be inverted to retrieve the corresponding values of $I_J$ through multiple iterations. This iterative process is needed to correctly take into account the scarrering contribution into the source function terms $S_i$.
During the first iteration of the Feautrier's routine, the scattering contribution has yet to be properly calculated. The source function at this step corresponds to the thermal blackbody radiation produced by the atmospheric layer $i$ at temperature $T_i$:
\begin{equation}
    S_i=B_i =B(T_i)
\end{equation}

For any other iteration, the model will consider the previous solution for $I_J$ to calculate the new source function, which will then include the contribution of the photons that have been scattered in the previous steps.
In the case of $i=N$, the source function must correspond to the right term in Equation \ref{eq:boundary}, computed using the most recent estimate of $I_+(P_{surf})$ and $J^{scat}(P_{surf})$.

This process can be accelerated though the Accelerated Lambda Iteration and Ng methods \citep[see][p. 75]{Molliere2017}).

From that value, it is possible to calculate $I_H$ using Equation \ref{eq:IJIH}. The emergent flux at the top of the atmosphere can be then calculated as follows:

\begin{equation} \begin{split}
    F&= \int_0^{2\pi} \int_0^{\pi/2} I(P=0)\cos{\theta}\sin{\theta}d\phi d\theta \\&= 2\pi \int_0^{\pi/2}I(P=0)\mu d\mu = -4\pi \int_0^1 I_H(P=0)\mu d\mu
    \end{split}
\end{equation}

The iterations stop once the estimate of the flux has reached a convergence value.

\section{Corner Plots}\label{app:corner}

\begin{figure*}
   \centering
   \includegraphics[width=0.95\linewidth]{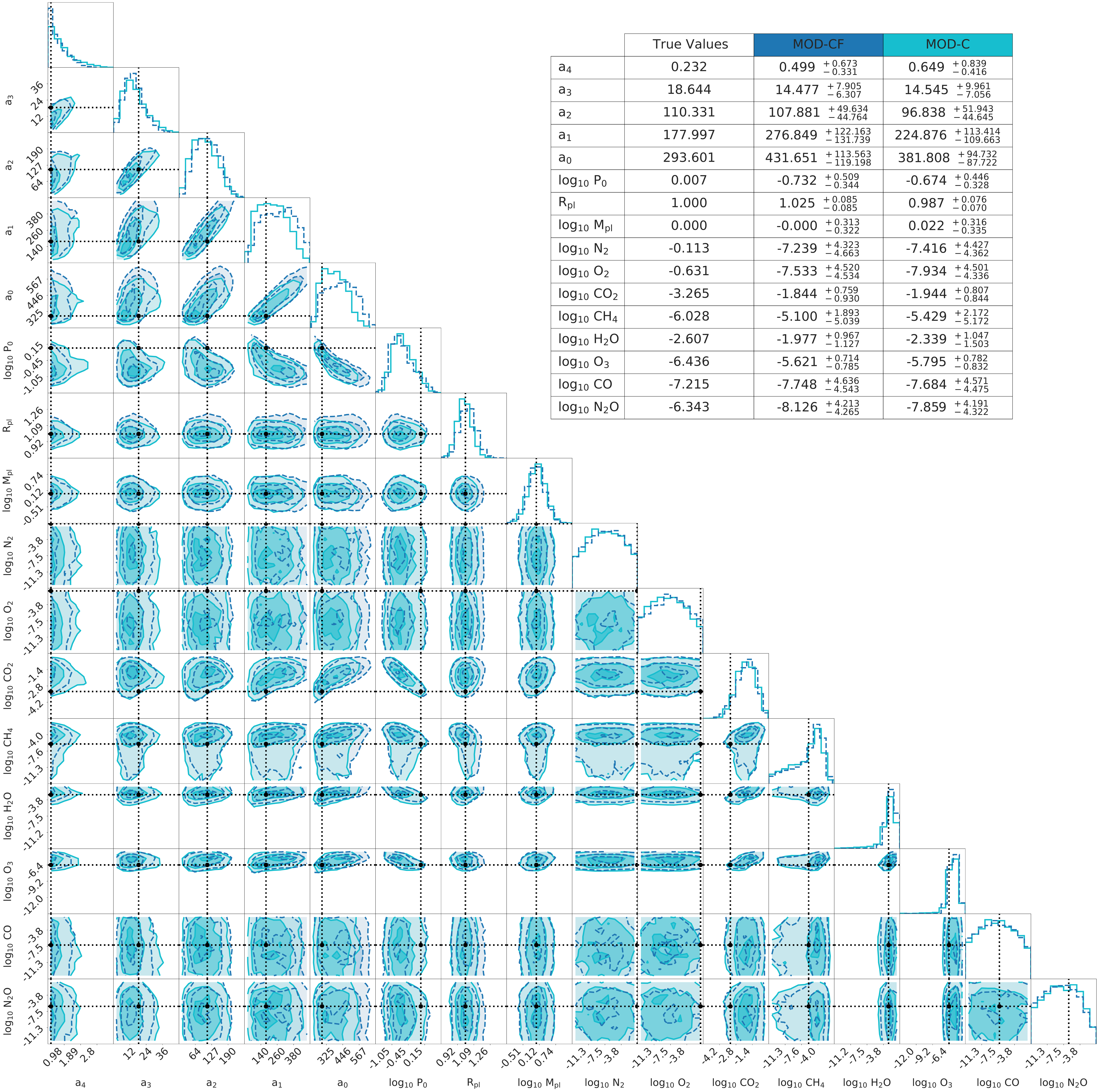}
      \caption{Corner plot for the posterior distributions from the retrievals of the clear Modern Earth (MOD-CF, dashed contour lines) and cloudy Modern Earth (MOD-C, solid contour lines) scenarios. The black
lines indicate the expected values for every parameter. The retrieved values (median and 1-$\sigma$ uncertainties) are shown in the Table in the top right corner, together with the expected values. The scenarios are color-coded according to Table \ref{table:id}.
      }
         \label{fig:corner_mod}
   \end{figure*}

\begin{figure*}
   \centering
   \includegraphics[width=0.95\linewidth]{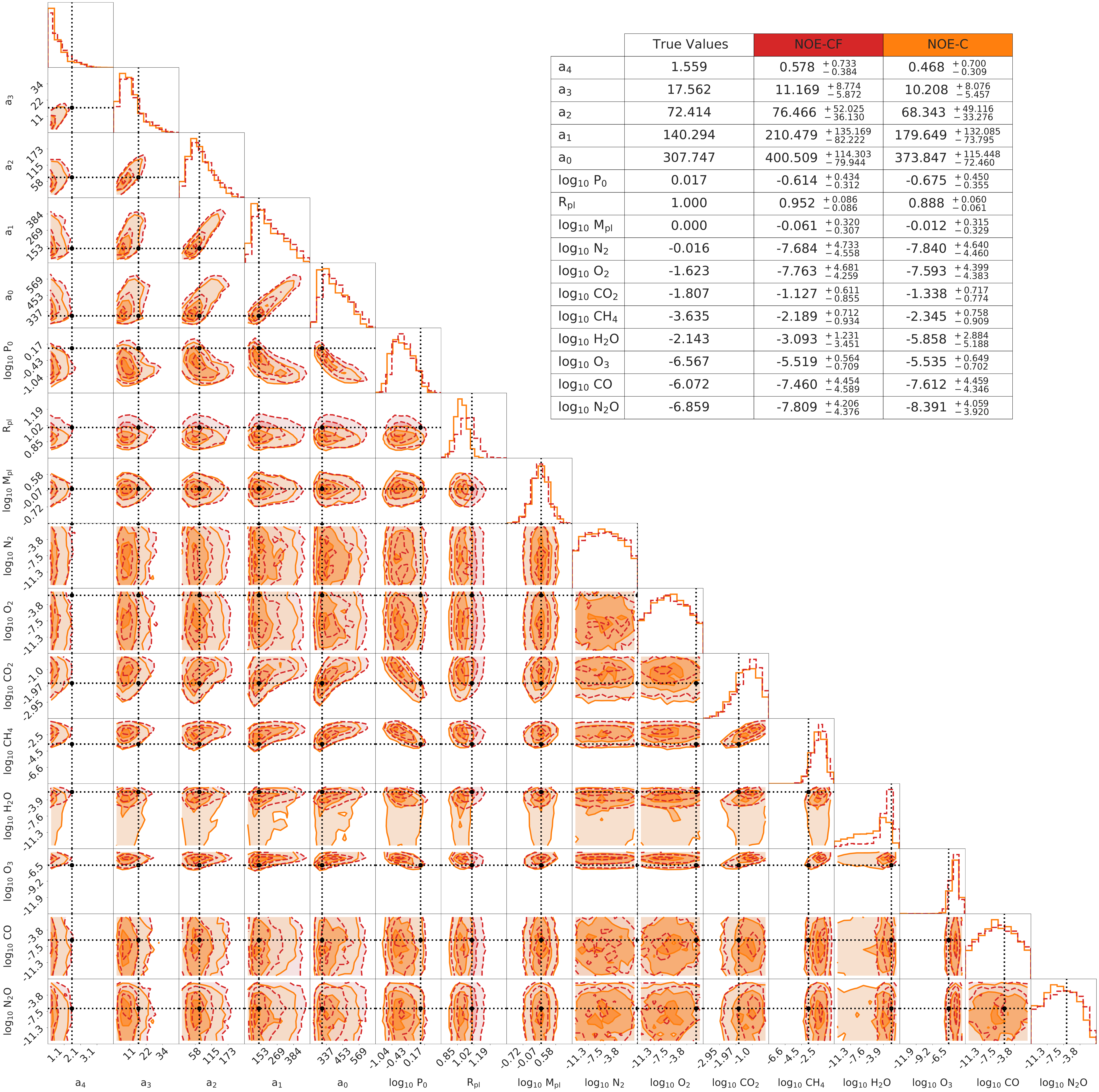}
      \caption{As for Figure~\ref{fig:corner_mod} but for the clear NOE Earth (NOE-CF) and cloudy NOE Earth (NOE-C) scenarios.}
         \label{fig:corner_noe}
   \end{figure*}

\begin{figure*}
   \centering
   \includegraphics[width=0.95\linewidth]{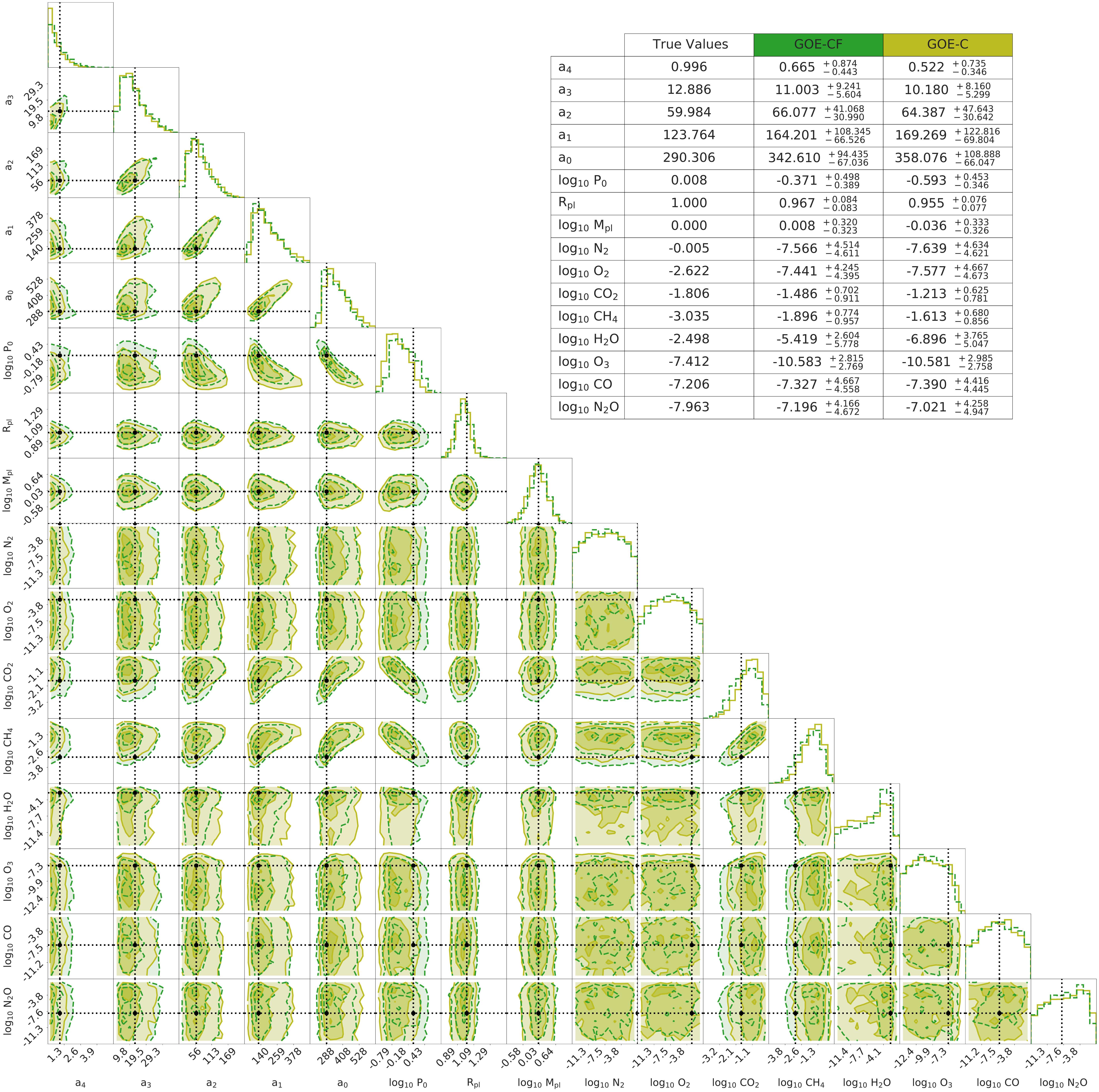}
      \caption{As for Figure~\ref{fig:corner_mod} but for the clear GOE Earth (GOE-CF) and cloudy GOE Earth (GOE-C) scenarios.}
         \label{fig:corner_goe}
   \end{figure*}

\begin{figure*}
   \centering
   \includegraphics[width=0.95\linewidth]{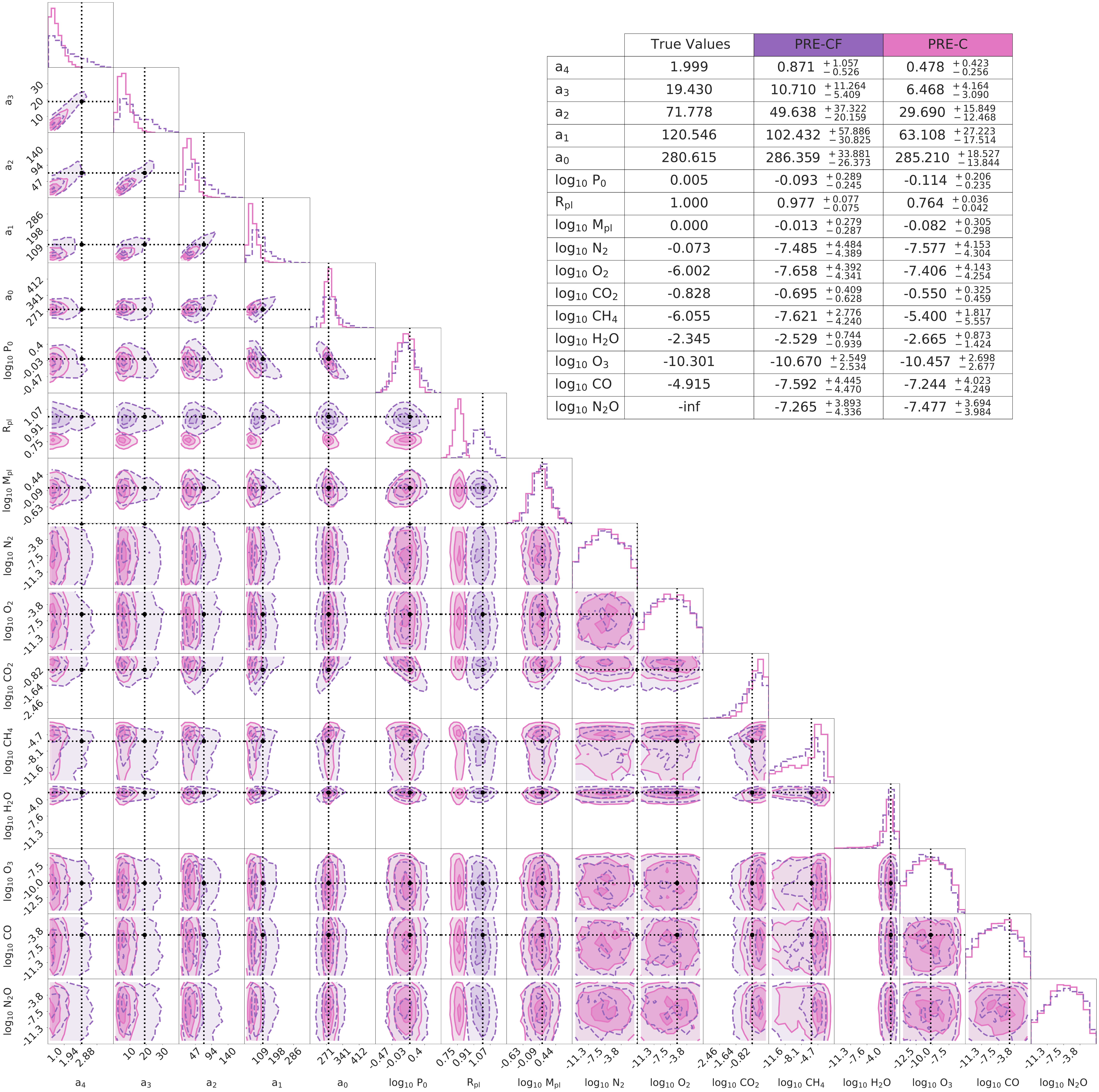}
      \caption{As for Figure~\ref{fig:corner_mod} but for the clear Prebiotic Earth (PRE-CF) and cloudy Prebiotic Earth (PRE-C) scenarios.}
         \label{fig:corner_pre}
   \end{figure*}

Corner plots for the retrieval runs at the reference R and S/N are shown in this section. We grouped both the cloudy and the clear sky retrievals for each epoch in the same figure, in order to compare the results. Namely: Figure~\ref{fig:corner_mod} shows the corner plots of the two Modern Earth scenarios (MOD-CF and MOD-C); Figure~\ref{fig:corner_noe} shows the NOE Earth scenarios (NOE-CF and NOE-C); the GOE Earth scenarios (GOE-CF and GOE-C) are in Figure~\ref{fig:corner_goe}; finally, the prebiotic scenarios (PRE-CF and PRE-C) are shown in Figure~\ref{fig:corner_pre}.

The models are color-coded according to Table~\ref{table:id}. Also, the results for the clear sky retrievals are shown using dashed contour lines, while the cloudy models are represented using solid lines.
The table on the top right of each figure shows the expected values for each parameter, together with the estimates and the 1-$\sigma$ uncertainty for the two scenarios. 

\section{Bayes' factor analysis: other epochs}\label{app:bayes}

As described in Section~\ref{sec:methods}, the theoretical spectral model was updated with respect to Paper III and it now takes into account additional physical processes. For the results presented in Section~\ref{sec:results}, we ran retrievals using the most updated version of the Bayesian framework. 


The additional flexibility of \texttt{petitRADTRANS} now allows us to quantify the impact of CIA and scattering in retrievals. We ran additional retrievals on the clear sky scenarios for R~$=50$ and S/N~$=10$. In these retrievals, we altered the number of physical processes that were treated in the \texttt{petitRADTRANS} theoretical spectral model as follows:
\begin{itemize}
    \item Including both CIA and scattering (setup used in Section~\ref{sec:results});
    \item Excluding both CIA and scattering;
    \item Including scattering and excluding CIA;
    \item Including CIA and excluding scattering.
\end{itemize}
In the runs where scattering is included, we consider self-scattering, surface scattering of the thermal radiation, and gaseous Rayleigh scattering (see Table~\ref{table:cia} for references). We do not include aerosol and cloud scattering in the calculation.
Since our theoretical spectral model neglects clouds, in this analysis we considered only the cloud-free scenarios. The effect of modeling cloudy spectra using a cloud-free retrieval model will be discussed in detail in the Section \ref{sec:clouds}.

To determine the theoretical spectral model configuration that best reproduces the input spectra we performed a Bayes' factor analysis. The Bayes' factor is defined as:
\begin{equation}\label{eq:bayes}
    K=\mathcal{Z}_{\mathcal{M}_1}(\mathcal{D})/\mathcal{Z}_{\mathcal{M}_2}(\mathcal{D})
\end{equation}
where $\mathcal{M}_1$ and $\mathcal{M}_2$ represent two different model configurations, each with their corresponding Bayesian evidence $\mathcal{Z}_{\mathcal{M}_i}(\mathcal{D})$ given the input data $\mathcal{D}$. In the case of Equation \ref{eq:bayes}, the Bayes' factor provides an indication whether $\mathcal{M}_1$ or $\mathcal{M}_2$ better describes the data. We can use the Jeffrey's scale (see Table~\ref{table:jeffrey}) to interpret the values of the Bayes' factor $K$.
This approach was extensively described and used in Paper III, to which we refer for more details.

\begin{figure*}
   \centering
    \includegraphics[width=\linewidth]{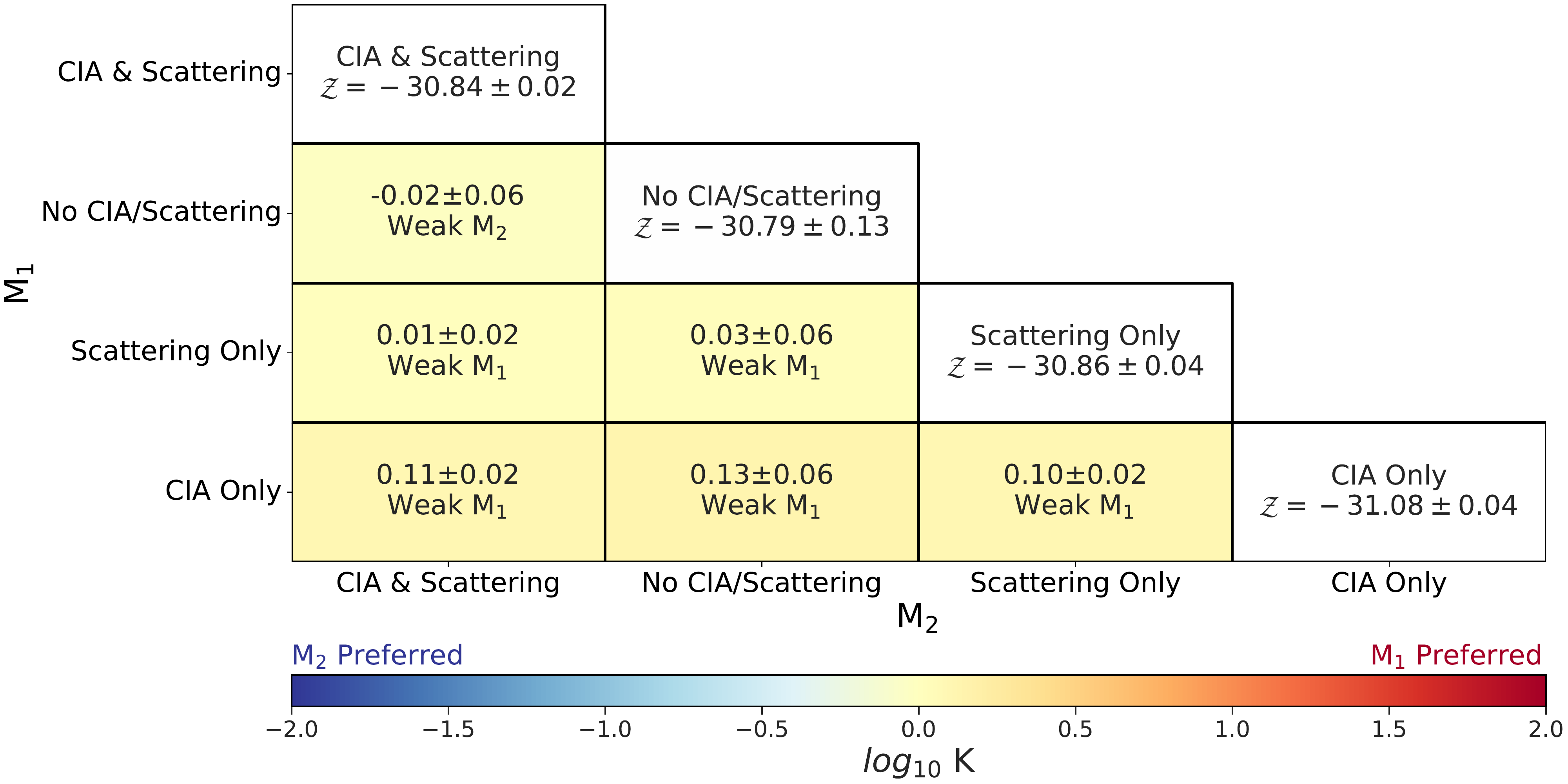}
   
   \caption{On the diagonal, the Bayesian evidence $\mathcal{Z}_{\mathcal{M}_i}$ for each setup for the clear Modern (MOD-CF) scenario. In the lower triangle, Bayes' factor for every pair of retrieval setups for the clear Modern Earth (MOD-CF) scenario. The cells in the lower triangle are color-coded according to the colorbar, whose limits are determined by the Jeffrey's scale (see Table~\ref{table:jeffrey}).  }
    \label{fig:ModernBayesian}
\end{figure*}

We calculate the Bayes' factor corresponding to every possible combination of the four different setups previously outlined. We summarize the results obtained for the clear Modern Earth (MOD-CF) epoch in Figure~\ref{fig:ModernBayesian}. 
Here, the diagonal shows the values of the Bayesian evidence $\mathcal{Z}_{\mathcal{M}_i}(\mathcal{D})$ of each of the four setups $\mathcal{M}_i$. The triangular matrix of 6 boxes below the diagonal is instead filled with the logarithm of the Bayes' factors $K$ (see Eq. \ref{eq:bayes}) for each combination of theoretical spectral model setups, as well as their interpretation according to the Jeffrey's scale (Table~\ref{table:jeffrey}). The cells are color-coded according to the color bar in the lower panel, whose edges are determined by the Jeffrey's scale.

\begin{table}
\caption{Jeffrey's scale \citep{Jeffreys:Theory_of_prob}.}
\label{table:jeffrey}      
\centering                          
\begin{tabular}{cl}        
\hline\hline                 
$\log_{10}\left(K\right)$ & Strength of Evidence\\    
\hline 
$(-\infty,-2]$              &Decisive support for $\mathcal{M}_2$\\
$(-2,-1]$            &Strong support for $\mathcal{M}_2$\\
$(-1,-0.5]$ &Substantial support for $\mathcal{M}_2$\\
$(-0.5,0]$          &Very weak support for $\mathcal{M}_2$\\
   $(0,0.5)$    &Very weak support for $\mathcal{M}_1$\\
   $[0.5,1)$   &Substantial support for $\mathcal{M}_1$\\
   $[1,2)$        &Strong support for $\mathcal{M}_1$\\
   $[2,\infty)$       &Decisive support for $\mathcal{M}_1$\\ 
\hline 
\end{tabular}
\tablefoot{Scale for the interpretation of the Bayes' factor $K=\mathcal{Z}_{\mathcal{M}_1}(\mathcal{D})/\mathcal{Z}_{\mathcal{M}_2}(\mathcal{D})$. Adapted from Paper III.}
\end{table}

The blue and red colour range adopted in the color bar were chosen deliberately in order to illustrate that, for any pair of models M$_1$ and M$_2$, a redder shade would mean that M$_1$ is more likely to reproduce the data compared to M$_2$, while a bluer shade would instead prefer M$_2$ over M$_1$. Our results show colors which lie somewhere in the middle of the range of the color bar, which corresponds to a value of $\log_{10}(K)$ generally very close to 0, as confirmed by the text within the cells.  This means that there is no clear preference for any of the tested setups: we find no evidence that one of the considered setups outperforms the others in describing the input data. Including or excluding CIA and scattering (one or both) results in negligible differences in the retrieval results. This means that CIA in the spectra or/and spectral features induced by scattering are
not detectable in retrieval studies at the considered R and S/N of the input. This analysis shows us that it is justifiable to neglect CIA and scattering in MIR retrievals of spectra with R~$=50$ and S/N~$=10$ (the minimum requirements for \emph{LIFE} determined by Paper III), with negligible loss in the quality of the retrieval results. 

\begin{figure*}
   \centering
    \includegraphics[width=\linewidth]{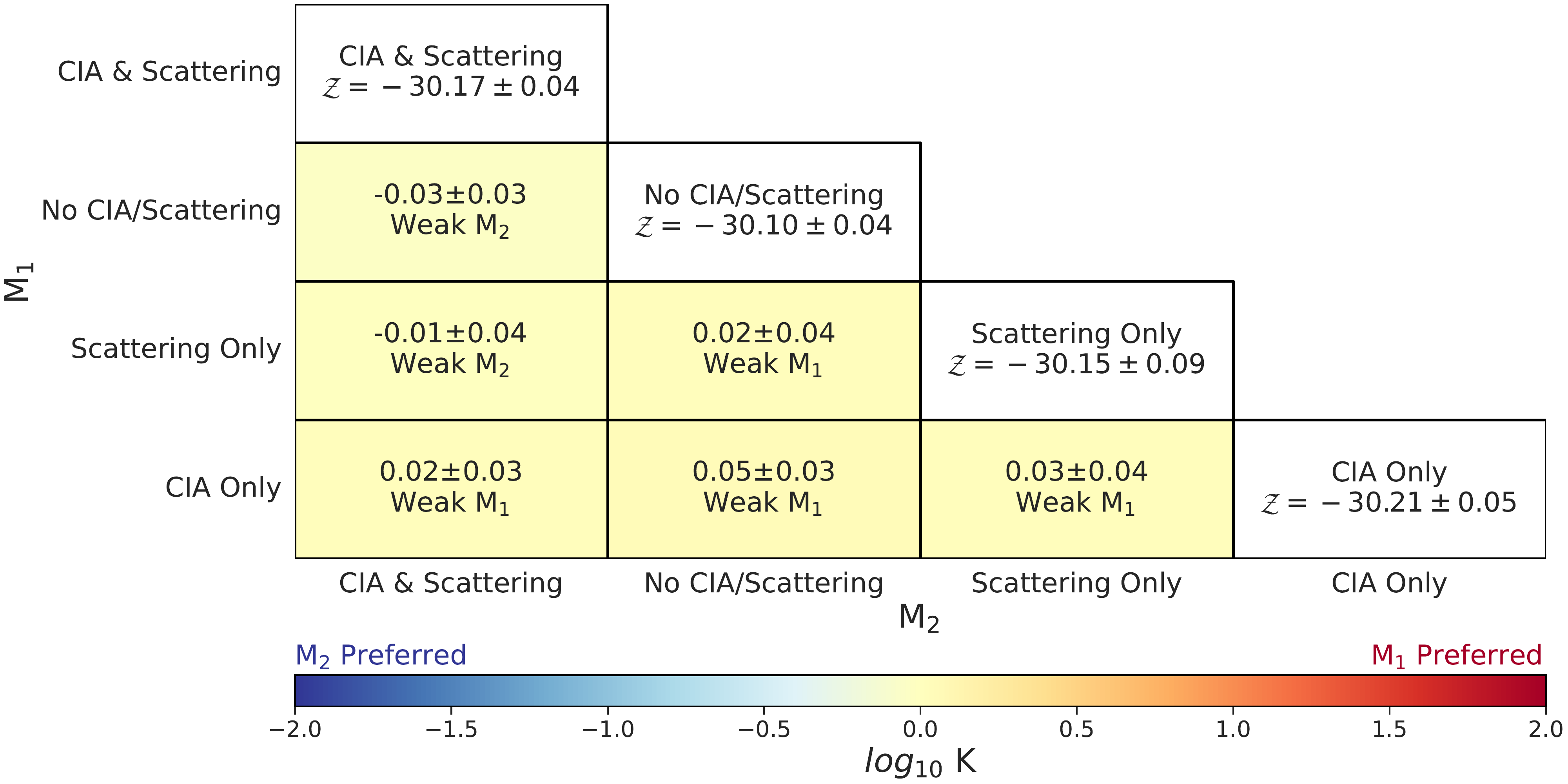}
   
   \caption{As for Figure~\ref{fig:ModernBayesian} but for the clear NOE Earth (NOE-CF) scenario.}
    \label{fig:NOEBayesian}
\end{figure*}

\begin{figure*}
   \centering
    \includegraphics[width=\linewidth]{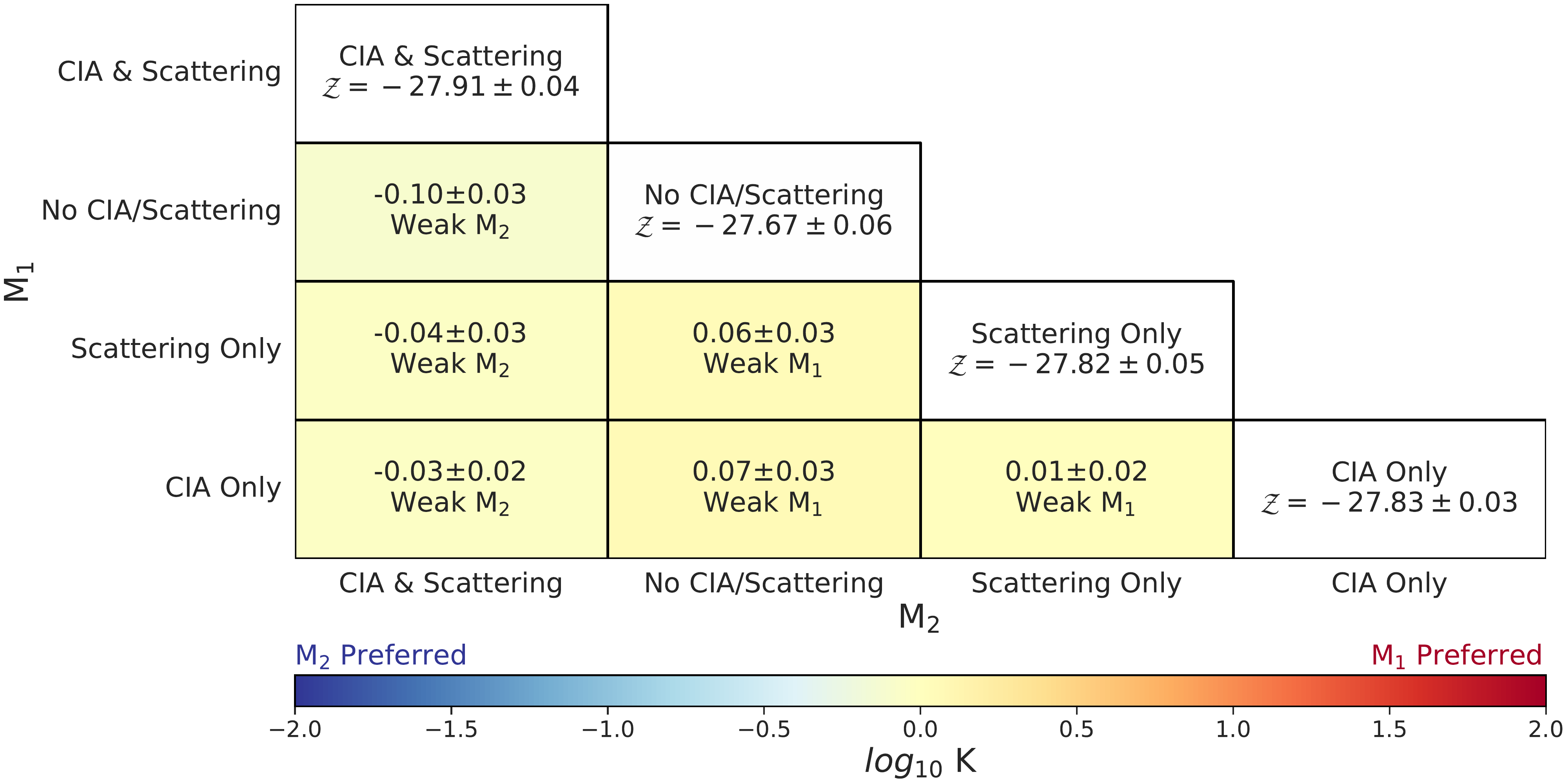}
   
   \caption{As for Figure~\ref{fig:ModernBayesian} but for the clear GOE Earth (GOE-CF) scenario.}
    \label{fig:GOEBayesian}
\end{figure*}

\begin{figure*}
   \centering
    \includegraphics[width=\linewidth]{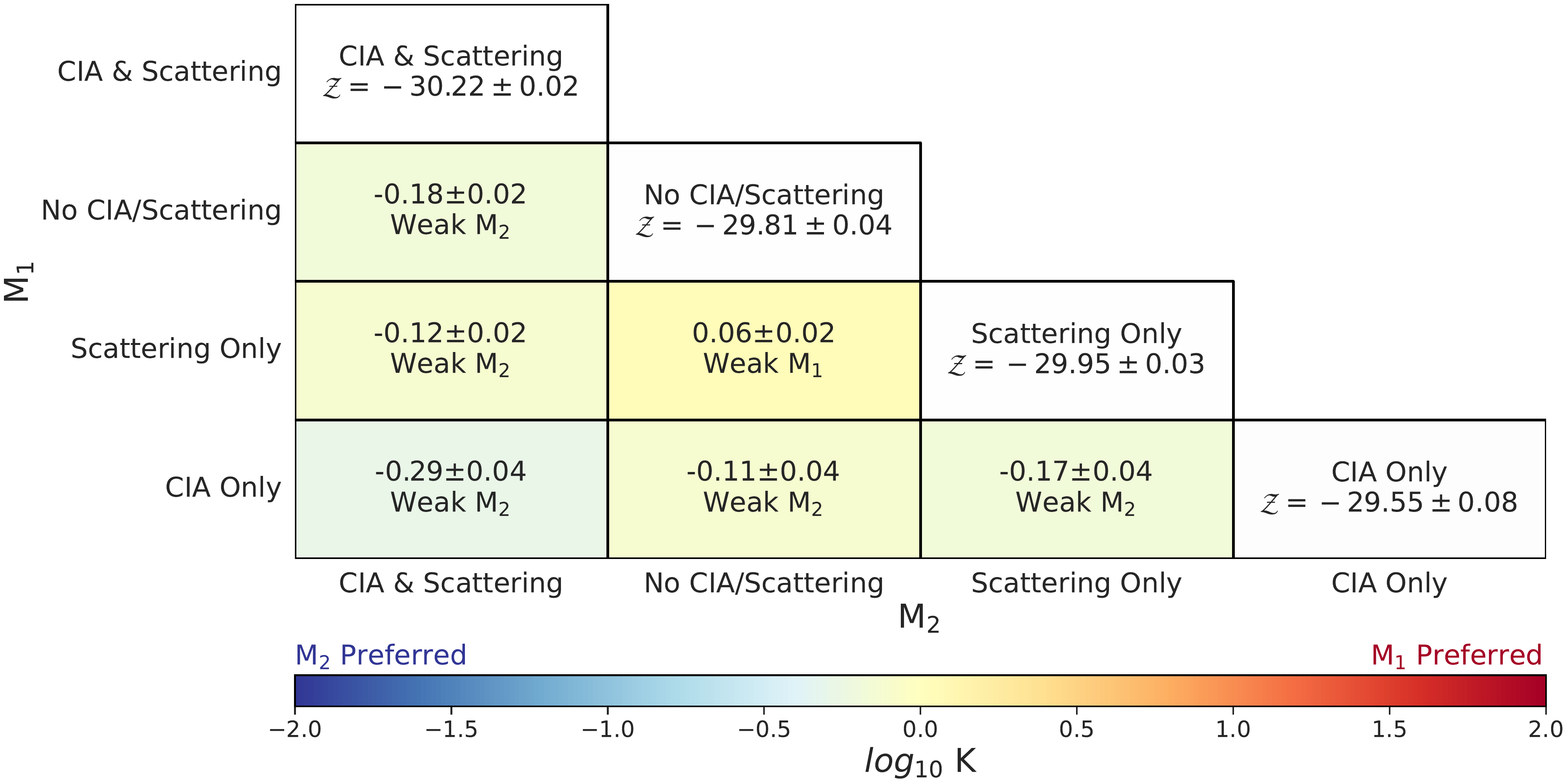}
   
   \caption{As for Figure~\ref{fig:ModernBayesian} but for the clear Prebiotic Earth (PRE-CF) scenario.}
    \label{fig:PrebioticBayesian}
\end{figure*}

The results for the remaining epochs exhibit similar behaviour. In Figure~\ref{fig:NOEBayesian}, we show the results for the NOE Earth, Figure~\ref{fig:GOEBayesian} shows the ones for the GOE Earth, and the results for the prebiotic Earth are shown in Figure~\ref{fig:PrebioticBayesian}. 

\section{Cloudy scenarios: additional figures}\label{app:cloudy}

\begin{figure}
   \centering
    \includegraphics[width=\linewidth]{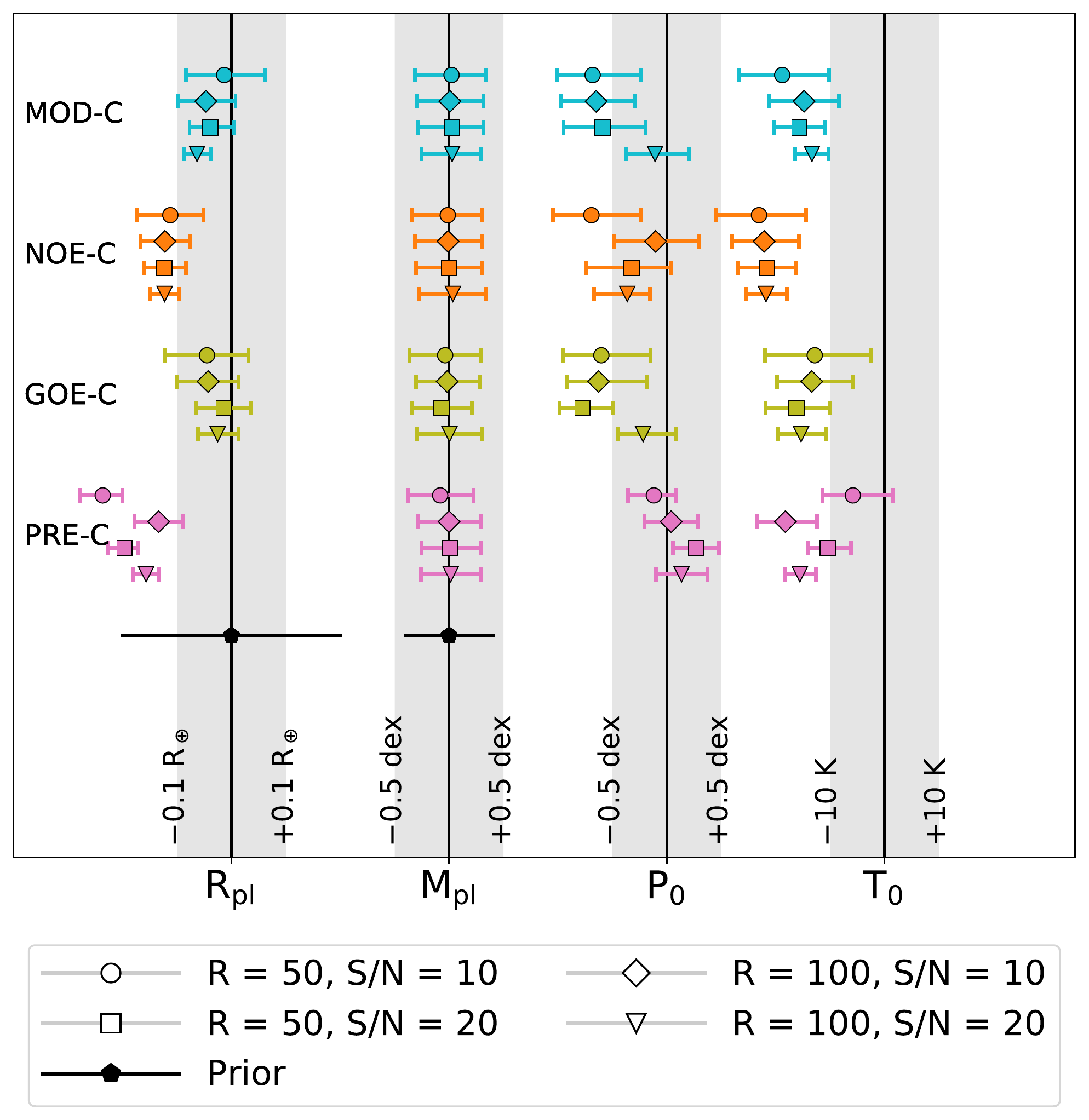}
   
   \caption{As for Figure~\ref{fig:QualityParams} but for the cloudy scenarios.}
    \label{fig:QualityParamsCloudy}
\end{figure}

\begin{figure*}%
    \centering
    \subfloat{{\includegraphics[width=.45\linewidth]{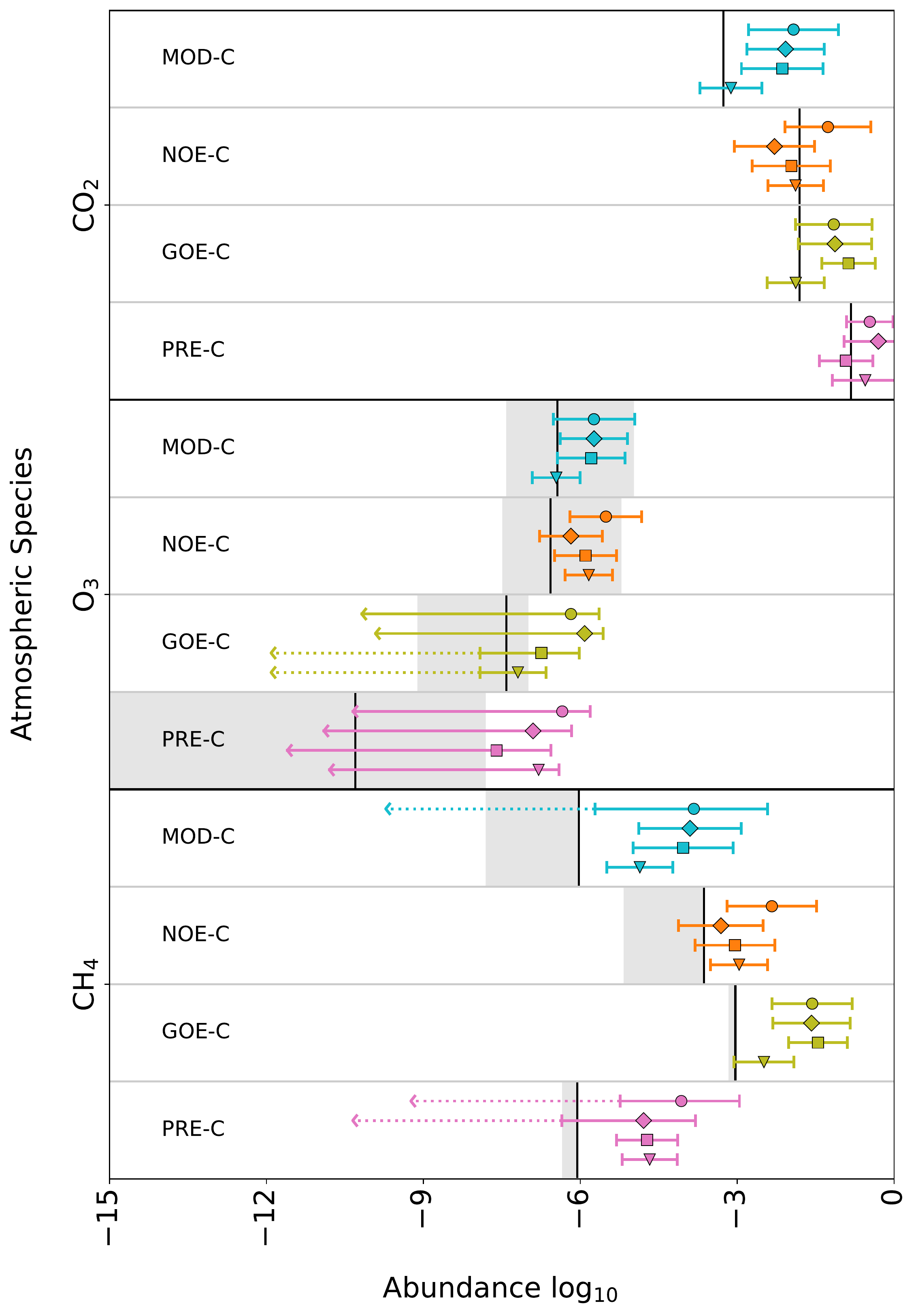} }}%
    \qquad
    \subfloat{{\includegraphics[width=.45\linewidth]{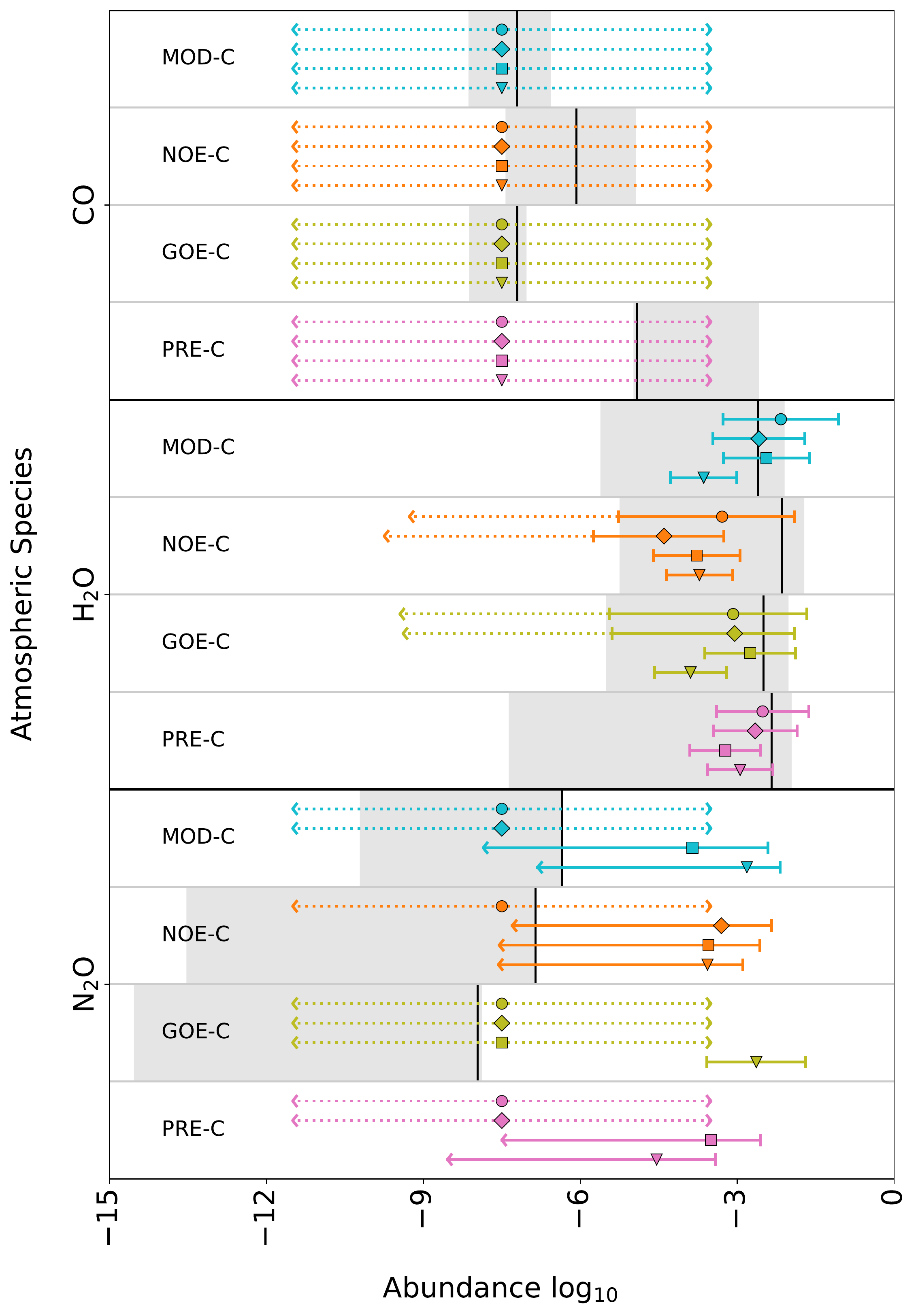} }}%
    \,
    \subfloat{\includegraphics[width=.4\textwidth]{images/abunds_legend.pdf}}
    \caption{As for Figure~\ref{fig:QualityAbunds} but for the cloudy scenarios.}
    \label{fig:QualityAbundsCloudy}%
\end{figure*}

\begin{figure*}
    \centering
    \includegraphics[width=\linewidth]{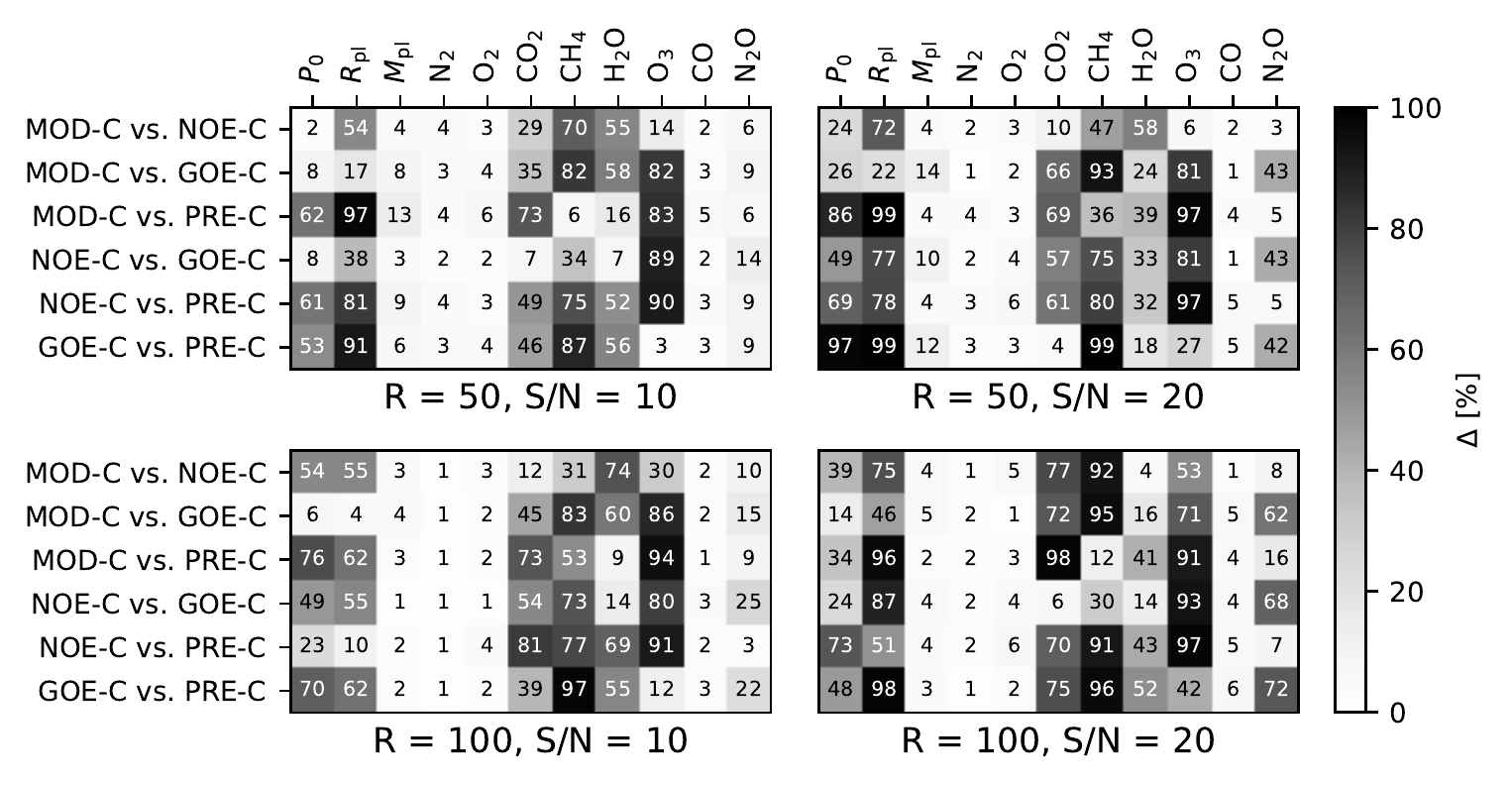}
    \caption{As for Figure~\ref{fig:DeltaAll}, but for the cloudy scenarios.}
    \label{fig:C_DeltaAll}
\end{figure*}

In this section, we provide additional plots for the cloudy scenarios. In Figures~ \ref{fig:QualityParamsCloudy} and \ref{fig:QualityAbundsCloudy} we show the retrieved exoplanet parameters and abundances for the different scenarios with varying R and S/N values. 
Finally, we plot in Figure~\ref{fig:C_DeltaAll} the maximum difference $\Delta$ between the cumulative posteriors for the different model parameters, for each combination of the cloudy scenarios and different R-S/N pairs.

\end{appendix}
\end{document}